\RequirePackage{lineno}
\RequirePackage{lineno}
 \documentclass[prd,twocolumn,preprintnumbers,superscriptaddress,nofootinbib]{revtex4}
\usepackage{graphicx}  % Include figure files
\usepackage{dcolumn}   % Align table columns on decimal point
\usepackage{bm}        % bold math
\usepackage{amssymb}   % for math
\usepackage{multirow}
\usepackage{lineno}
\usepackage[hyperfootnotes=false]{hyperref}
\begin{document}
%%%%**********LINE NUMBERS COMMENT OUT
%\linenumbers
%%%%% -----
%\doublespace
%%%%%%%%%%%%%%%%%%%%%%%%%%%%%%%%%%%%%%%%%%%%%%%%%%%%%%%%%%

\newcommand{\carbon}{\rm ^{12}C}
\newcommand{\oxygen}{\rm ^{16}O}
\newcommand{\deuteron}{\rm ^{2}H}
\newcommand{\hydrogen}{\rm ^{1}H}
\newcommand{\Hefour}{\rm ^{4}He}
\newcommand{\lead}{\rm^{208}Pb}
\newcommand{\Hethree}{\rm ^{3}He}
\newcommand{\neon}{\rm ^{20}Ne}
\newcommand{\aluminum}{\rm^ {27}Al}
\newcommand{\argon}{\rm ^{40}Ar}
\newcommand{\iron}{\rm ^{56}Fe}
\newcommand{\genie}{$\textsc{genie}$}
\newcommand{\gep}{$G_{Ep}$ } 
\newcommand{\gen}{$G_{En}$ } 
\newcommand{\qv}{$\bf |\vec q|$}
\newcommand{\rlqe}{$R_L^{QE}(\bf q, \nu)$ }
\newcommand{\rtqe}{$R_T^{QE}(\bf q, \nu)$ }
\newcommand{\rltot}{$R_L(\bf q, \nu)$ }
\newcommand{\rttot}{$R_T(\bf q, \nu)$ }
\newcommand{\csr} {$S_L({\bf q})$}
\newcommand{\Rochester}{Department of Physics and Astronomy, University of Rochester, Rochester, NY  14627, USA}
\newcommand{\JLAB}{Thomas Jefferson National Accelerator Facility, Newport News, VA 23606, USA}
%%%%%%%%%%%%%%%%%%%%%%%%%%%%%%%%%%%%%%%%%%%%%%%%%%%%%%%%%%
%
% \hspace{5.2in} \mbox{PRD letter  v5 Aug 25, 2022}
%\hypersetup{
%    pdftitle={\title{Extraction of the  Coulomb  Sum Rule 
 % for ${\rm ^{12}C}$  and $^{16}$O 
%  } },
%    pdfauthor={A. Bodek}}
\title{Contribution of Nuclear Excitation Electromagnetic Form Factors in  ${\rm ^{12}C}$  and ${\rm ^{16}O}$ to the Coulomb Sum Rule} 
%%%%%%%%%%
%\input{authors_inputs}
\affiliation{\Rochester}
\affiliation{\JLAB}
  \author{A.~Bodek}
%   \thanks{Corresponding author  Arie Bodek, bodek@pas.rochester.edu}
\affiliation{\Rochester}
    \author{M.~E.~Christy}
\affiliation{\JLAB}
 %   christy@jlab.org
 % M. Eric Christy
%\input{authors_list}
\date{\today} 
%3:25 pm
\begin{abstract}
We report on empirical parameterizations of longitudinal and transverse nuclear excitation electromagnetic form factors in ${\rm ^{12}C}$ and ${\rm ^{16}O}$. We extract the contribution of  nuclear excitations to the Normalized Inelastic Coulomb Sum Rule (\csr) as a function of momentum transfer $\bf q$ and find that it is significant (0.29$\pm$0.030 at $\bf q$= 0.22 GeV). The total contributions of nuclear excitations to $S_L({\bf q})$ in ${\rm ^{12}C}$ and ${\rm ^{16}O}$ are found to be equal within the uncertainties.  Since the cross sections for nuclear excitations are significant, the radiative tails from nuclear excitations should be included in precise calculations of radiative corrections to quasielastic electron scattering at low $\bf q$ and deep-inelastic electron scattering at large energy transfers $\nu$. The parameterizations also serve as a benchmark in testing theoretical modeling of cross sections for excitation of nuclear states in  electron and neutrino interactions on nuclear targets  at low energies.
\end{abstract}
\maketitle  
\section{Introduction}
\label{intro}
The Normalized Inelastic Coulomb Sum Rule \csr~\cite{CSR} in electron scattering on nuclear targets  is the integral of the longitudinal  nuclear response function   $R_L({\bf q},\nu)d\nu$ ({\it excluding the nuclear elastic peak and  pion production processes}) divided by the square of the proton electric form factor and by
 the number of protons in the nucleus.  Here, $\bf q$ is the momentum transfer and $\nu$ is the energy transfer to the nuclear target. The sum rule has contributions from quasielastic  (QE) scattering and from the electro-excitations of nuclear states. At high $\bf q$ it is expected that $S_L \rightarrow 1$ because both nuclear excitation form factors and Pauli suppression are small. At small $\bf q$ it is expected that $S_L \rightarrow 0$ because all cross sections for inelastic processes (e.g. QE,  nuclear excitation and pion production processes) must be zero at $\bf q$=0. 

In this paper we present details of empirical parameterizations  of the $\bf q$ dependence of all  longitudinal and transverse excitation form factors in ${\rm ^{12}C}$. Since there are fewer measurements on ${\rm ^{16}O}$ we only parameterize the longitudinal form factors for this nucleus.  We use these parameterizations  to compute the contribution of nuclear excitations to $S_L({\bf q})$ for both nuclei.  Our investigation of the  QE contribution to $S_L({\bf q})$ is reported in an earlier publication\cite{short_letter}. 
 
Since the cross sections for nuclear excitations are significant at low ${\bf q}$, the parametrizations should be used in precise calculations of radiative corrections to quasielastic electron scattering  at low ${\bf q}$.  Because of intial state radiation, nuclear excitations also contribute to radiative corrections in deep-inelastic  electron scattering at large  $\nu$. 
 The parameterizations also serve as benchmark in testing theoretical modeling of electron and neutrino scattering at low energies. 
 Because of recent advances in theoretical methods\cite{Lovato2016,microscopic,Coupled} for  the calculations of the response functions of electron scattering on nuclear targets, it is now possible to make theoretical predictions of the  form factors for the excitation of nuclear states in both electron and neutrino scattering\cite{Pandey1,Pandey2,Pandey3}.
 %
%
 %Figure 1    FFFFFFFFFFFFFFFFFFFFFFFFFFFFFFFFFFFFFFFFFFF
%
\begin{figure}[ht]
\begin{center}
\includegraphics[width=3.4in,height=1.7in]{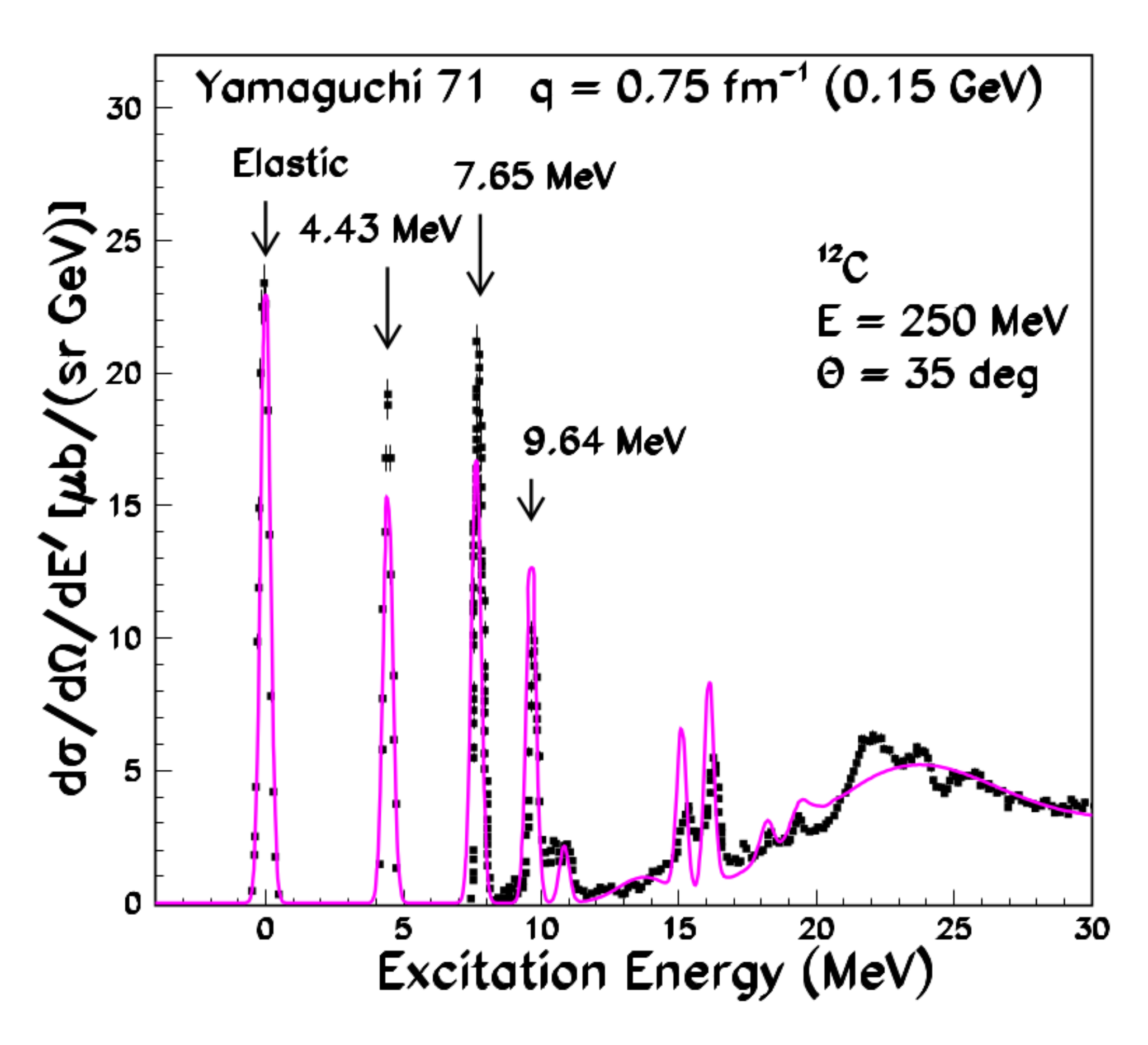}
\includegraphics[width=3.4in,height=1.7in]{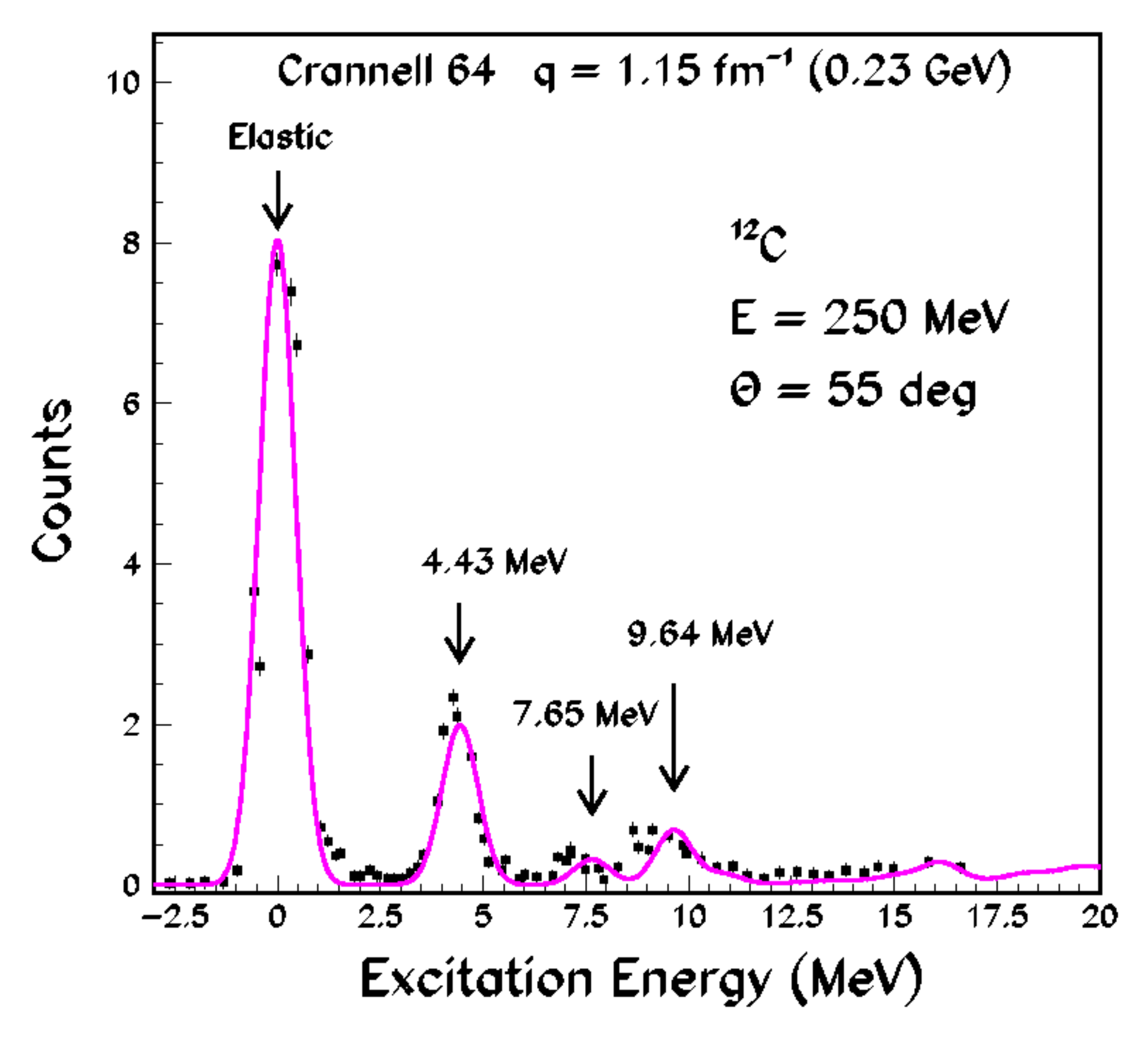}
\includegraphics[width=3.4in,height=1.7in]{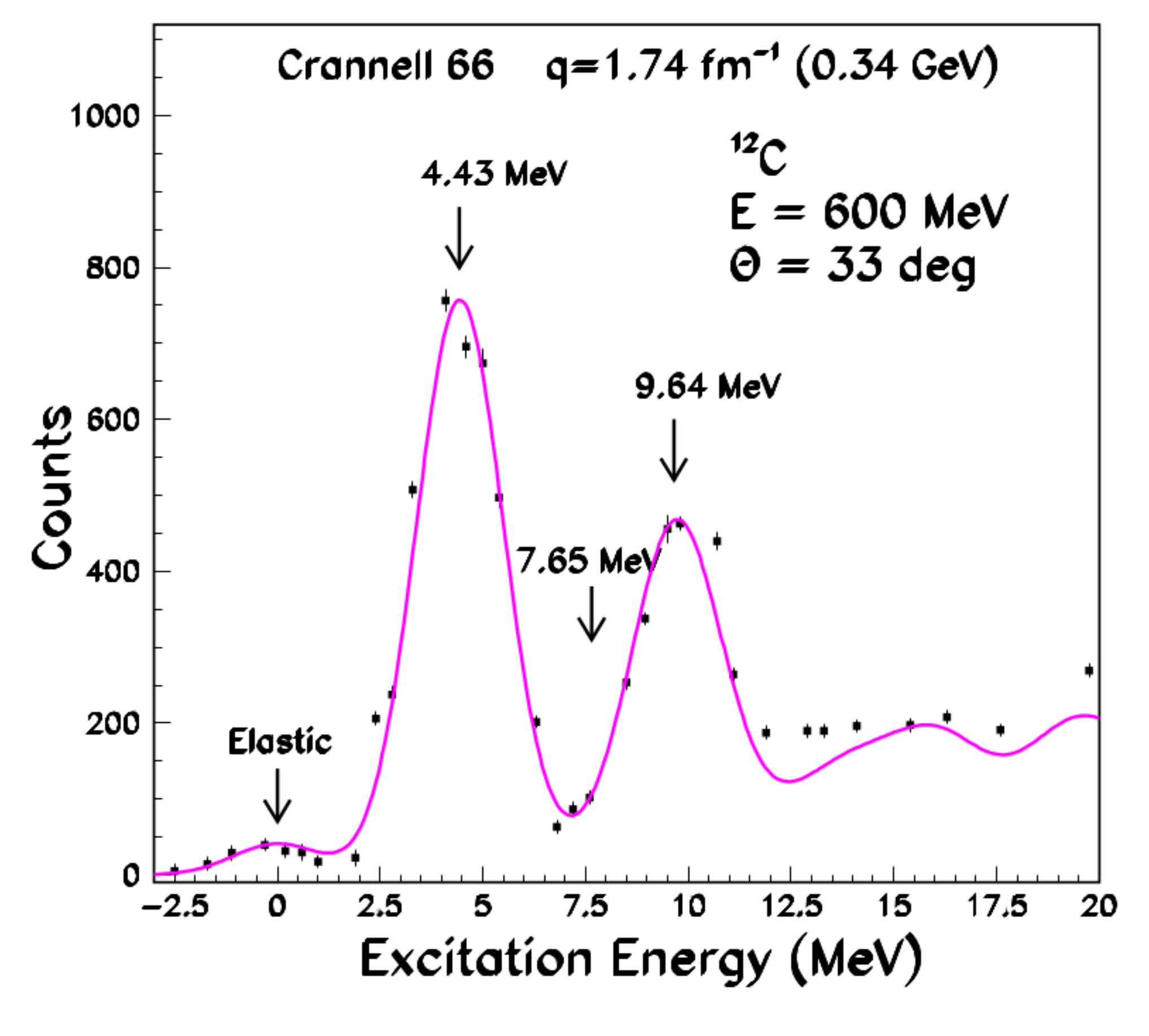}
\caption{{\bf Top panel}: Radiatively corrected cross section from Yamaguchi\cite{Yamaguchi}(measured with high resolution of 0.25\%) for the scattering of 250 MeV electrons from  ${\rm ^{12}C}$ at 35$^0$.  Here, the cross section for the elastic peak has been divided by 100 and the cross section for the 4.43 MeV state by 4.    
%(${\bf q}_{elastic}$ = 0.75 fm$^{-1}$).
 {\bf Middle panel}: Radiatively corrected cross section\cite{Crannell1}  (in arbitrary units)  for the scattering of 250 MeV electrons from  ${\rm ^{12}C}$ at 55$^0$.
  %(${\bf q}_{elastic}$=1.18 fm$^{-1}$). 
   {\bf Bottom panel}:  Radiatively corrected cross section\cite{Crannell1}  (in arbitrary units)  for the scattering of 600 MeV electrons from ${\rm ^{12}C}$ at 33$^0$. 
   %(${\bf q}_{elastic}$ =1.75 fm$^{-1}$).  
   The peaks for elastic scattering and for  the first three nuclear excitations at 4.43, 7.66 and  9.64 MeV are clearly visible. The solid curve is the predicted radiatively corrected cross section using our fits to the form factors and QE cross sections. The fit is  normalized to the elastic cross section for the  E=250 MeV and  55$^0$ data. For the E= 600 MeV and 33$^0$ data we normalize to the cross section for the  4.43 MeV state.}
\label{C12_states_1}
\end{center}
\end{figure} 
%     FFFFFFFFFFFFFFFFFFFFFFFFFFFFFFFFFFFFFFFFFFF

\begin{figure}
\begin{center}
\includegraphics[width=3.4in,height=4.7in]{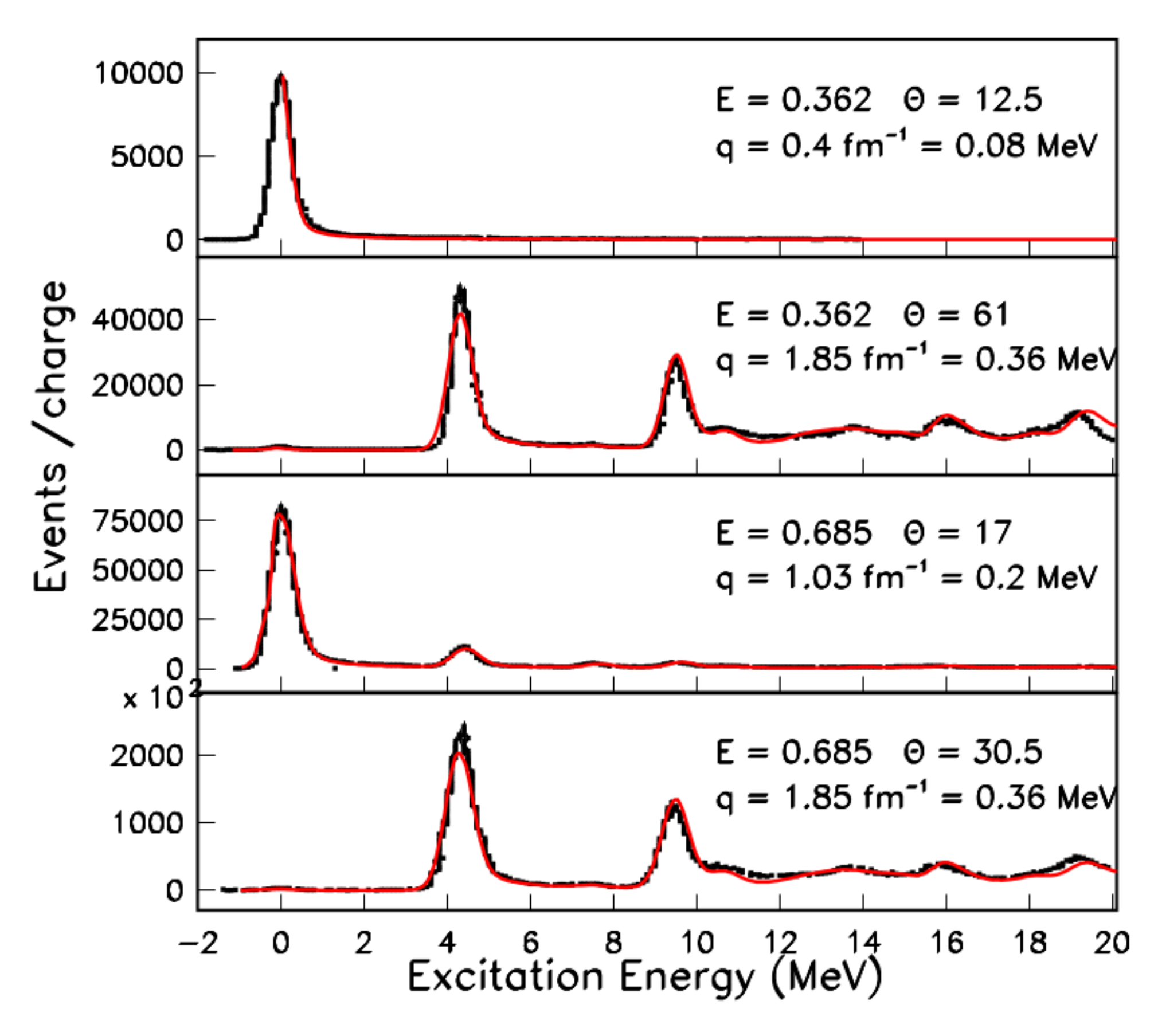}
\caption {Radiatively {\it uncorrected} cross section (in arbitrary units) from the LEDEX experiment\cite{ledex_exp} on ${\rm ^{12}C}$.
% Here,  from top to bottom panels,   ${\bf q}_{elastic}$ are 0.4 fm$^{-1}$, 1.85 fm$^{-1}$, 1.03 fm$^{-1}$, and 1.82 fm$^{-1}$.
   The solid red line is the  radiatively  {\it uncorrected} cross section from  our fit  to the form factors and QE cross sections. It is normalized to the elastic cross section at zero excitation energy for the  12.5 and 17.0 degree data, and to the cross section for the 4.43 MeV state for the 30.5 and 61.0 degree data).
 }
\label{LEDEX_fig}
\end{center}
\end{figure}

 % Fig 1 and Fig 2
 Figures  \ref{C12_states_1}  and  \ref{LEDEX_fig} show the relative contributions of the cross sections for  elastic scattering from the ${\rm ^{12}C}$ nucleus, as well as the low lying  excitations of nuclear states for several low energy data sets~\cite{Crannell1,Yamaguchi, ledex_exp}.  Also shown as a solid curve is our parameterization utilizing the experimental resolution to apply a Gaussian smearing to each state.
 % for a few incident energies and scattering angles. 
%For ${\rm ^{12}C}$ we  parameterize the $\bf q$ dependence of both the longitudinal and  transverse excitation %form factors. For  ${\rm ^{16}O}$ we  only parameterize the longitudinal form factors because there are fewer %published measurements,   
%
%
%SSSSSSSS NEW   SECTION
\section{Theoretical framework}
The electron scattering differential cross section can be written in terms of longitudinal ($R_L({\bf q},\nu)$) and
transverse ($R_T({\bf q},\nu)$) nuclear response functions \cite{GEp}:
%
%  Eq. 1 
\begin{eqnarray}
\frac{d\sigma}{d \nu d\Omega}&=& \sigma_M [ A R_L ({\bf q},\nu) + B R_T ({\bf q}, \nu)]
\end{eqnarray} where  $\sigma_M$ is the Mott cross section, 
%
%  eq  2
\begin{eqnarray}
 \sigma_M &=&\frac {\alpha^2 \cos^2)(\theta/2)}{4 E_0^2 \sin^4(\theta/2)}.
\end{eqnarray}
Here,  $E_0$ is the incident electron energy,  $E^{\prime}$ is energy of the final state electron, 
$\nu=E_0-E^{\prime}$ is the energy transfer to the target, $\bf q$ is the 3-momentum transfer, $Q^2$ is the square of the 4-momentum transfer (defined to be positive such that ${\bf q}^2= Q^2+\nu^2$), $A=(Q^2/{\bf q}^2)^2$ and $B = \tan^2(\theta/2) +Q^2/2{\bf q}^2$.  For nuclear elastic scattering at very low $\bf q$ on ${\rm ^{12}C}$ $Q^2 = {\bf q}^2$ to a good approximation.
 %
%Fig. 2

For elastic scattering and nuclear excitations the square of the electric and magnetic form factors are obtained by the integration of the measured response functions over $\nu$.  In the  experimental extractions of form factors for elastic scattering and nuclear excitations the Mott cross section is defined with an additional factor of $Z^2$ because both the nuclear elastic cross section and the  cross sections for the the electro excitation of nuclear  states are proportional to $Z^2$ times  charge form factors $F_{iC}^2 ({\bf q})$. Here, the subscript zero denotes the nuclear elastic peak and subscripts 1-N denote nuclear excitations.  The charge form factors can be thought of as the product\cite{ledex_exp,product} of the proton electric form factor and the  form factors of the spatial  distribution of protons in the nucleus.

\section {Coulomb Sum Rule}
The inelastic Coulomb Sum Rule is the integral of  $R_L({\bf q},\nu)d\nu$, {\it excluding the elastic peak and  pion production processes}. It has contributions from QE scattering and from electro-excitations of nuclear states: 
%\vskip -1 mm
\begin{eqnarray}
\label{CSReq}
&&{\rm CSR}({\bf q})= \int R_L({\bf q},\nu)d\nu\\
&=& \int R_L^{QE}({\bf q},\nu) d\nu\nonumber
%&=& \int \rlqe d\nu 
+ G^{\prime 2}_E(Q^2) \times Z^2 \sum_{all}^{L} F^2_i({\bf q})\\\nonumber
&=& G^{\prime 2}_E(Q^2) \times \big[ Z \int V^{QE}_L({\bf q},\nu) d\nu + Z^2 \sum_{all}^{L} F^2_i({\bf q})\big]\nonumber.
\end{eqnarray}
%
%\rlqe is the longitudinal QE response and 
We define $V^{QE}({\bf q},\nu)$ as the reduced longitudinal QE response, which integrates to unity in the absence of any suppression (e.g. Pauli blocking).
The  charge form factors for the electro-excitation of nuclear states $F^2_{iC}({\bf q})$ is related to $F^2_i({\bf q})$ by the expression  $ F^2_{iC}({\bf q}) = G^{\prime 2}_{Ep} (Q^2)\times F^2_i({\bf q})$. 
%are the product of the spatial distribution form factor $F^2_i({\bf q})$ times the electric form factor of the proton $G_{EP}^\prime ({\bf q})$
In order to account for the  small contribution of the neutron and relativistic effects $G^{\prime 2}_E(Q^2)$ is given by\cite{GEp}:
 \begin{equation}
 G^{\prime 2}_E(Q^2)= [G^2_{Ep}(Q^2)+\frac{N}{Z}G^2_{En}(Q^2)] \frac{1+\tau}{1+2\tau},
 \label{GEprime}
  \end{equation}
where, \gep and \gen are the electric form factors~\cite{BBBA} of the proton and neutron, respectively and  $\tau=Q^2/4M_p^2$. 
%We use the BBBA07 parametrizations~\cite{BBBA} of the free  nucleon form factors. 
%At low $Q^2$ the quantity in brackets in equation \ref{GEprime} is well represented by a simple dipole ($1/(1+Q^2/0.71)^4$). 
%
%  G^{\prime 2}_E(Q^2)
By dividing Eq.~\ref{CSReq} by $ZG^{\prime 2}_E {\bf q})$ we obtain the normalized inelastic Coulomb Sum Rule $S_L({\bf q})$ :
\begin{eqnarray}
\label{SLINE}
S_L({\bf q})= \int V^{QE}_L({\bf q},\nu)d\nu + Z \sum_{all}^{L} F^2_i({\bf q}).
\end{eqnarray}

%At high $\bf q$ it is expected that $S_L \rightarrow 1$ because both nuclear excitation form factors and Pauli suppression are small. At small $\bf q$ it is expected that $S_L \rightarrow 0$ because the all form factors for inelastic processes (QE and nuclear excitations) must be zero at $\bf q$=0.  

%SSSSS  NEW SECTION
\section {Parameterization of ${\rm ^{12}C}$ nuclear elastic and nuclear excitation form factors}
\label{section_C12_states}
  %
%
%Fig 3.
% FFFFFFFFFFFFFFFFFFFFFFFFFFFFFFFFFFFFFFFFFFF
\begin{figure}
\begin{center}
\includegraphics[width=3.5in,height=4.in]{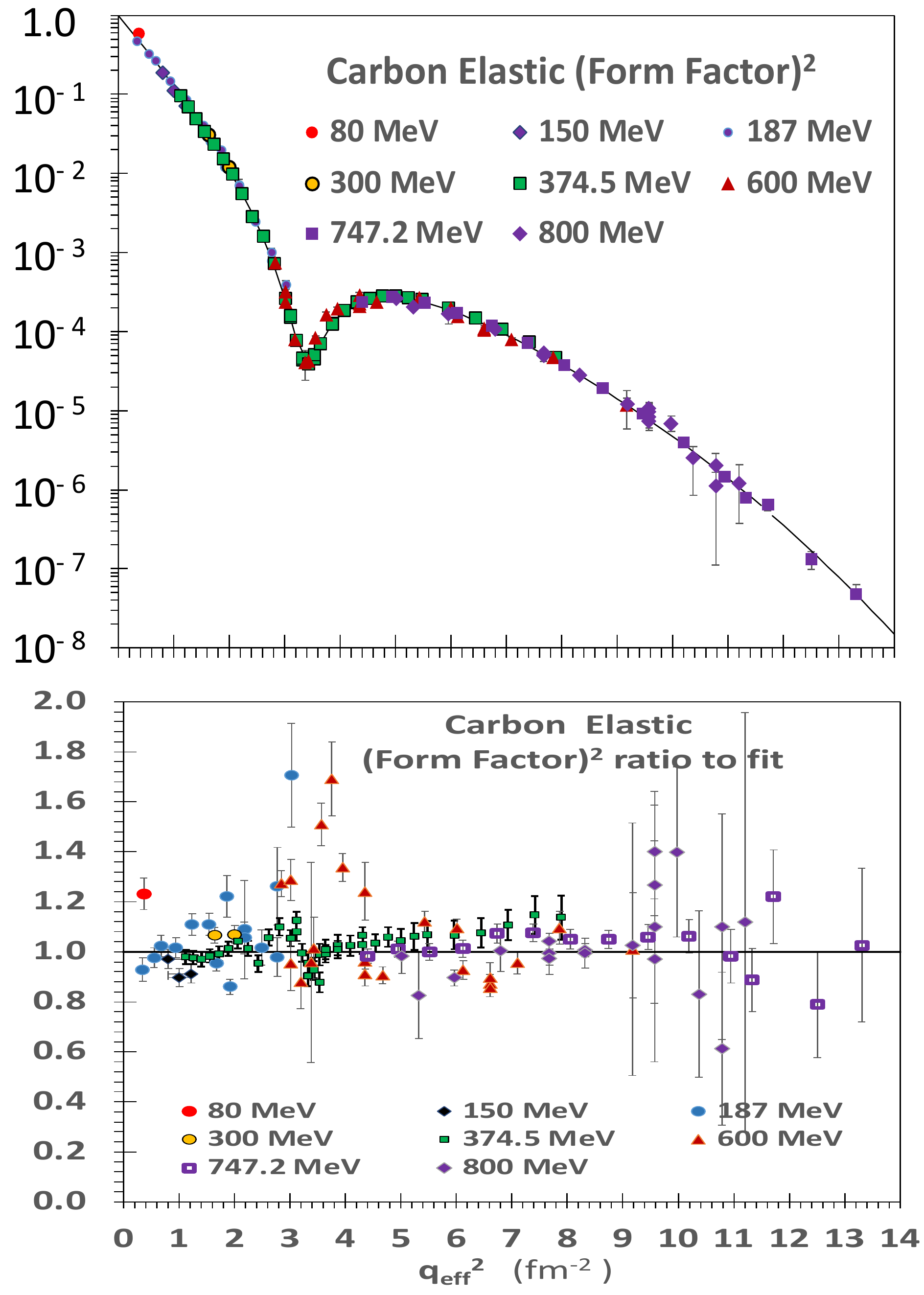}
\caption{{\bf Top panel}: Measurements\cite{C12_FF} of the nuclear elastic  longitudinal charge form factor (squared) for ${\rm ^{12}C}$ versus $\bf q^2_{eff}$. {\bf  Bottom panel}: Ratio to our fit.}
\label{C12elastic}
\end{center}
\end{figure} 
%FFFFFFFFFFFFFFFFFFFFFFFFFFFFFFFFFFFFFFFFFFF

% Figure 4 
%FFFFFFFFFFFFFFFFFFFFFFFFFFFFFFFFFFFFFFFFFFF
\begin{figure*}
\begin{center}
\includegraphics[width=3.5in,height=4.1in]{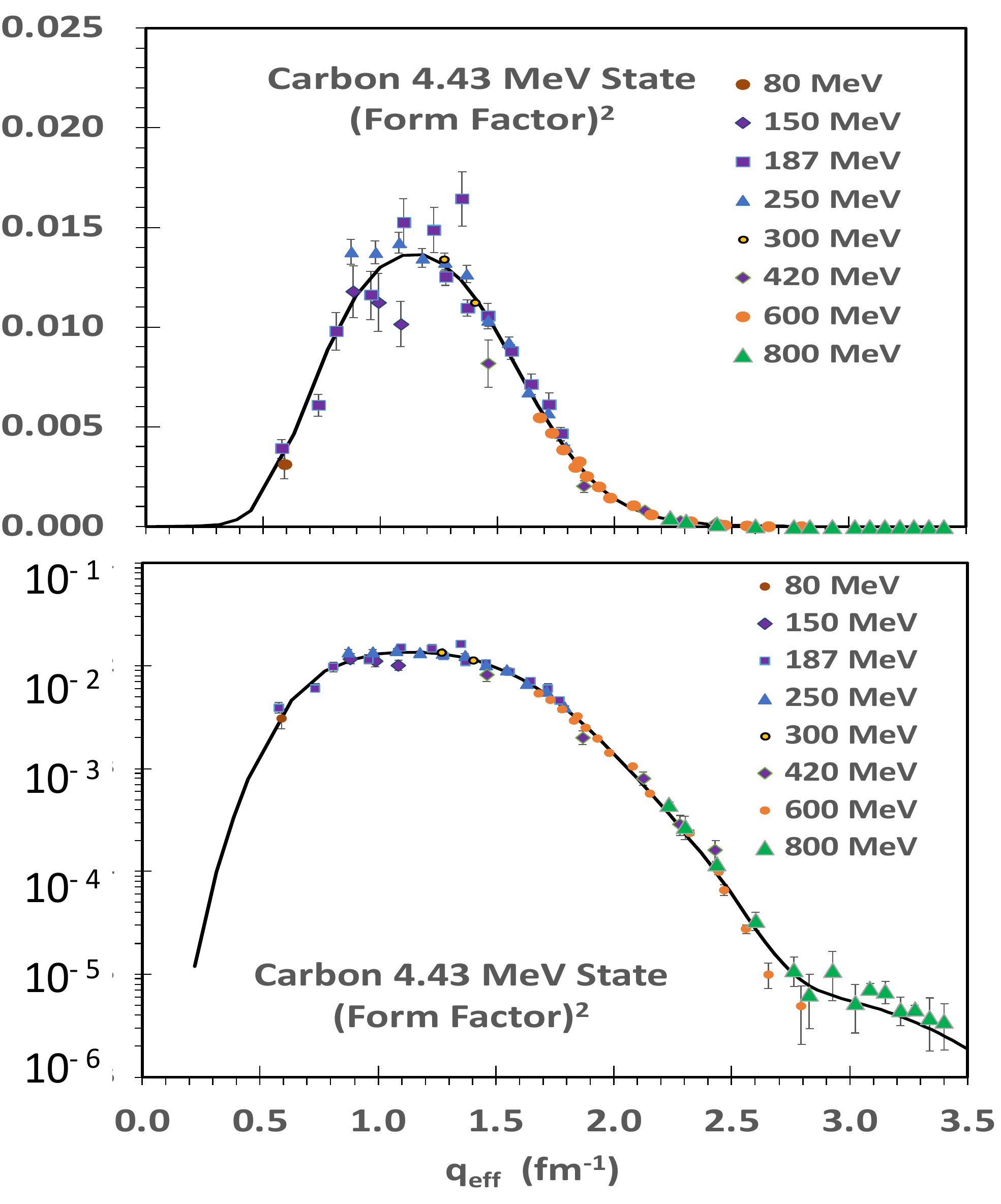}
\includegraphics[width=3.5in,height=4.05in]{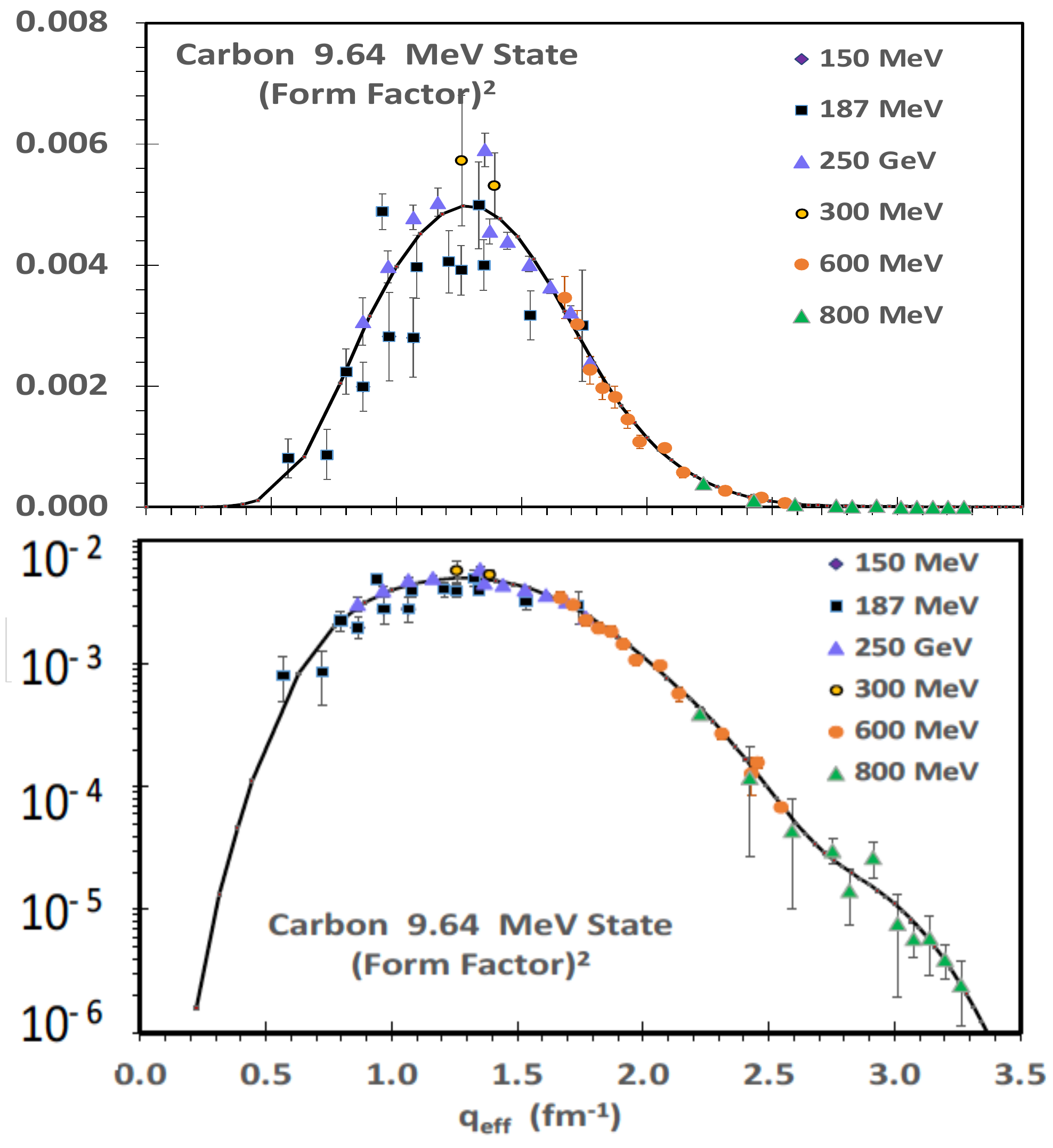}
\caption{ Measurements\cite{C12_FF} of the longitudinal charge form factors (squared) for the 4.43 MeV state (left) and the 9.64 MeV state (right) in ${\rm ^{12}C}$. The  form factors (squared)
are shown on linear scales and logarithmic scales on the top and bottom panels, respectively.}
\label{C12states}
\end{center}
\end{figure*} 
%FFFFFFFFFFFFFFFFFFFFFFFFFFFFFFFFFFFFFFFFFFF

%Figure  5 FFFFFFFFFFFFFFFFFFFFFFFFFFFFFFFFFFFFFFFFFFF
\begin{figure*}
\begin{center}
\includegraphics[width=3.5in,height=2.7in]{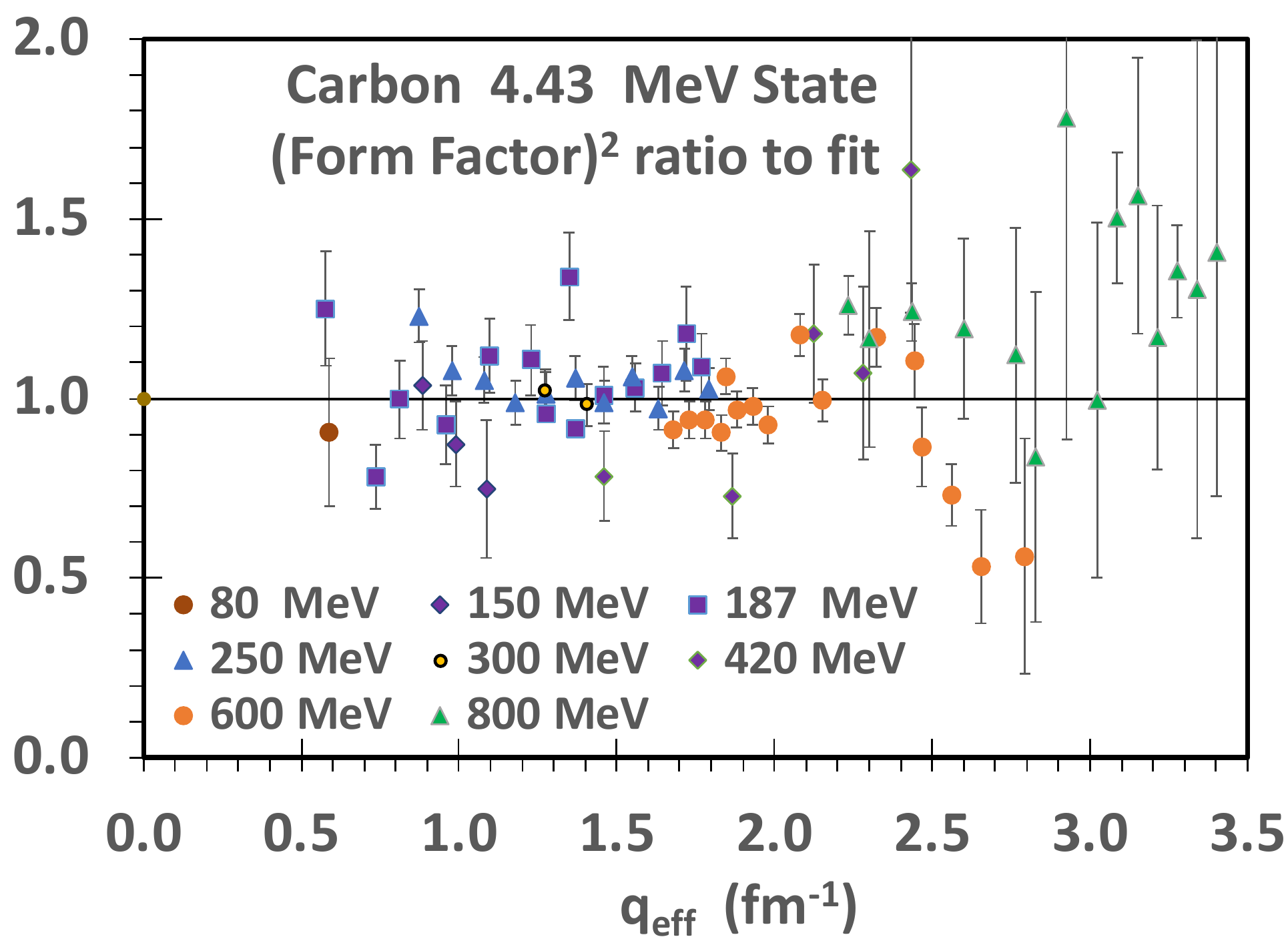}
\includegraphics[width=3.5in,height=2.7in]{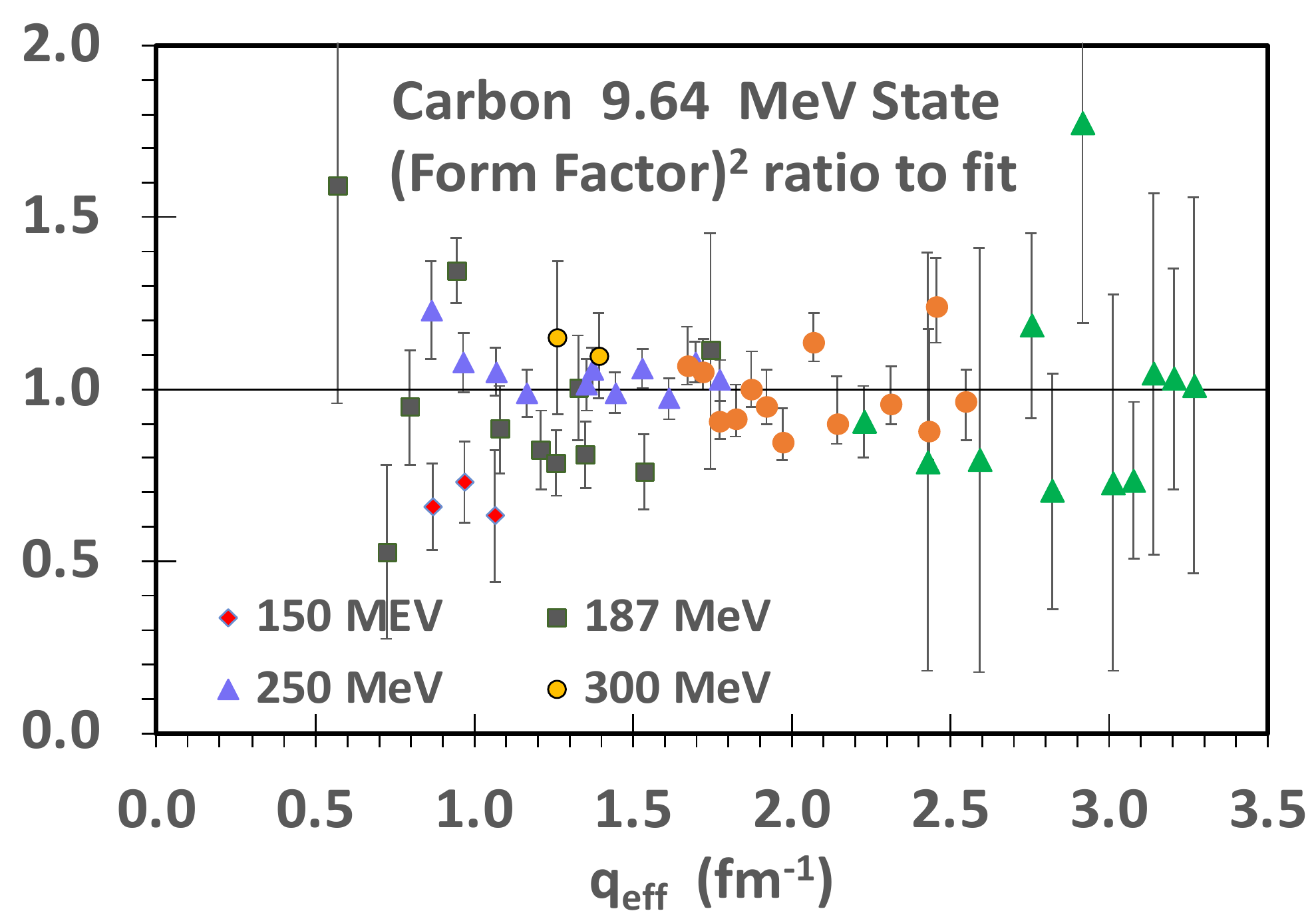}
\caption{Ratios of the measured\cite{C12_FF} longitudinal charge form factors (squared) to our parametrizations for the  4.43 MeV state (left) and the 9.64 MeV state (right) in ${\rm ^{12}C}$.}
 \label{C12_ratios}
\end{center}
\end{figure*} 
%FFFFFFFFFFFFFFFFFFFFFFFFFFFFFFFFFFFFFFFFFFF
%%
 \subsection { ${\rm ^{12}C}$ elastic form factor versus  ${\bf q}^2_{\rm eff}$}
 %%%
 The ${\rm ^{12}C}$ nucleus has a spin parity of $0^+$. We  fit the measured ${\rm ^{12}C}$ elastic longitudinal (charge) form  factor with the following functional form:
 %  Eq. 23
   \begin{equation}
 F_{oC}^2({\bf q}^2_{\rm eff})=\frac{1+1.5\times 10^{-3}{\bf q}^4_{\rm eff}}{1+e^{Power}}[H^2( {\bf q}^2_{\rm eff})+G({\bf q}^2_{\rm eff})]
   \end{equation}
 Here,  $Power=\frac{ {\bf q}^2_{\rm eff}-12.0}{1.4}$ is included to better describe the form factor at very large ${\bf q}$.  The effective\cite{q_effective}   ${\bf q}^2$ is  
   $${\bf q}^2_{\rm eff}={\bf q}^2 (1+4Z\alpha/(3\langle r^2 \rangle E)$$
  Which for carbon is
   ${\bf q}^2_{\rm eff}={\bf q}^2 (1+ 0.00465/E)^2$ (where E is in GeV).
 The function  $H({\bf q_{\rm eff}}^2)$ is the  harmonic well shape with ($\alpha$= 1.21, and $a_0$=1.65).
  It is is given by\cite{Hofstadter}:
  %  Eq 24
 \begin{equation}
H({\bf q}^2_{\rm eff}) =[1-\frac{\alpha {\bf q}^2_{\rm eff} a_0^2}{2(2+3\alpha)}] exp[\frac{{\bf -q}^2_{\rm eff}a_0^2}{4}],
\end{equation}
The function $G({\bf q}^2_{\rm eff})$   fills in the dip in the location of the diffraction minimum. 
 $$  G({\bf q}^2_{\rm eff})= 5.0 \times 10^{-5} e^{-[({\bf q}^2_{\rm eff} -3.1)/0.66)]^2}$$
In the above parametrization ${\bf q}^2_{\rm eff}$ is  in units of fm$^{-2}$.
 A comparison of the parametrization of the nuclear elastic charge form factor for~ ${\rm ^{12}C}$ to experimental data\cite{C12_FF}  is shown on the top panel of Fig.  \ref{C12elastic}. The ratio of the measurments to the fit is shown on the bottom panel.
 %

%
%% Table 3 TTTTTTTTTTTTTTTTTTTTTTTTTTTTTTTTTTTTTTTTTTTTTTTTTTTTTTTTTTT  UPDATED
 \begin{table*}
  %[ht]
\begin{center}
\begin{tabular}{|c||c|c|c||c|c|c||c|c|c||c||c|} \hline
State	&	$N_1$	&	$C_1$	&	$\sigma_1$		&	$N_2$ &	$C_2$	&	$\sigma_2$	&	$N_3$ &	$C_3$	&	$\sigma_3$	 & $d$ & Data from Ref. \\ \hline \hline
%
%\bf{4.44 MeV-L} ($\bf q^2$) &	$1.43\times 10^{-2}$	&	1.050	&	1.70		&	$7.2\times 10^{-4}$	 &	3.00	&	2.0	&	$7.0 \times 10^{-6}$ &	7.0	&	5.0	 & 0.07 \\ \hline 
%
\bf{4.44 MeV~2$^+$ L} ($\bf q^2_{eff}$) &	$1.41\times 10^{-2}$	&	1.125	&	1.71	&	$7.2\times 10^{-4}$	 &	3.00	&	2.0	&	$7.0 \times 10^{-6}$ &	7.6	&	5.0	 & 0.10 &  \cite{C12_FF} \\ \hline \hline
%
%\bf{9.64 MeV-L} ($\bf q^2$)&	$5.00\times 10^{-3}$	&	1.40	&	1.70		&	$6.6\times 10^{-4}$	 &	3.30	&	1.9	&	$2.1 \times 10^{-5}$ &	7.0	&	2.5	 & 0.18  \\  \hline
%($\bf q^2_{eff}$)
\bf{9.64 MeV~3$^-$L} ($\bf q^2_{eff}$) &	$5.00\times 10^{-3}$	&	1.46	&	1.70		&	$6.6\times 10^{-4}$	 &	3.46	&	1.9	&	$2.1 \times 10^{-5}$ &	7.0	&	2.5	 & 0.20 &  \cite{C12_FF}  \\  \hline
%%  UPDATED
%  -  Dec 6, 33  d for 4.4  0.07-->0.11  and for 9.6 0.18-->0.20
 \hline
 \end{tabular}
 \end{center}
\caption{ Parameters of our fits (eq.  \ref{verses_q2}) to the ${\rm ^{12}C}$  longitudinal charge form factors (squared) for the 4.44 and 9.64 MeV nuclear excited states in ${\rm ^{12}C}$. For these states, the parametrizations are in terms of  $\bf q_{\rm eff} ^2$   in units of fm$^{-2}$. 
Here   ${\bf q^2_{\rm eff}}={\bf q^2} \times (1+ 0.00465/E)^2$, where E is in GeV\cite{q_effective}.}
\label{excited_states1}
\end{table*} 
%   TTTTTTTTTTTTTTTTTTTTTTTTTTTTTTTTTTTTTTTTTTTTTTTTTTTTTTTTTTT
%
% Table  4  TTTTTTTTTTTTTTTTTTTTTTTTTTTTTTTTTTTTTTTTTTTTTTTTTTTTTTTTTTT
\begin{table*}[ht]
\begin{center}
\begin{tabular}{|c||c|c|c||c|c|c||c|c|c||c|c||c|} \hline
State	&	$N_1$	&	$C_1$	&	$\sigma_1$		&	$N_2$ &	$C_2$	&	$\sigma_2$	&	$N_3$ &	$C_3$	&	$\sigma_3$	& $a $ &$b$ & Ref.\\ \hline \hline
{\bf 7.65 MeV 0$^+$L}&	$2.8\times 10^{-3}$	&	0.93	&	0.42		&	$3.0\times 10^{-4}$	&	1.45	&	0.24	&	$2.0 \times 10^{-5}$ &	2.48	&	0.53	& $1.0 \times 10^{-4}$ & 1 &  \cite{FF_765} \\ \hline \hline
{\bf 10.84 MeV 1$^-$L} &	{5.0 $\times 10^{-4}$}	&	1.0	&	0.3		&	{$8.0\times 10^{-4}$} &	{1.4}	&	0.4	&	$-$ &	-	&	-	& - & -  &
\cite{FF_1084}\\ \hline \hline
11.83 MeV 2$^-$ T &		{ 3.9 $\times 10^{-5}$}	&	1.2	&	0.5		&	{$1.2\times 10^{-5}$} &	{2.0}	&	0.4	&	- &	-	&	-	& - &-&\cite{Hicks84}  \\  \hline \hline
12.71 MeV 1$^+$T &	{ 3.0 $\times 10^{-6}$}	&	0.63	&	0.4		&	{$1.0\times 10^{-8}$} &	{1.0}	&	0.1	&	$2.0 \times 10^{-6}$ &	{ 1.8}	&	0.6	& $2.5\times 10^{-5}$ & 1  & \cite{FF_1271_1511}\\ \hline \hline
{\bf 13.7 MeV  4$^-$L} &{4.0 $\times 10^{-4}$}	&	1.0	&	0.35		&	 {$1.0\times 10^{-3}$}  &	1.75	&	0.45 &	 {$4.0\times 10^{-4}$}	&0.85	& 0.65 & -& &
 \cite{Crannell1,ledex_exp,Yamaguchi} \\
 $\sigma$=1.25 MeV 	&		&	&			&		&		&		&	&		&		&  & -& \\  \hline \hline
{\bf 14.08 MeV 4$^+$L} &{2.4 $\times 10^{-5}$}	&	1.8	&	0.6		&	- &	-	&	- &	-	&	-	& - & -& &
 \cite{FF_1408} \\ \hline \hline
%-
{\bf 15.1 MeV 1$^+$L}&	{$6.0\times 10^{-4}$}	      & 0.85	&	0.7 	      &	-	&	-	&	-	&	- &	-	&	-	& - & -& \cite{FF_1271_1511}  \\ 
15.1 MeV-T &	{ 2.5 $\times 10^{-4}$}	&	0.63	&	0.4		&	{$2.8\times 10^{-4}$} &	{ 0.84}	&	0.2	&	$2.4 \times 10^{-5}$ &	{ 2.0}	&	0.5	& $2.5\times 10^{-5}$ & 1 &  \cite{FF_1271_1511}\\  \hline \hline
%e{yamaguchi
\bf {16.1 MeV 2$^+$L}&	{\bf $12.0\times 10^{-4}$}	&1.05	&	0.6		&	-	&	-	&	-	&	- &	-	&	-	& - & - &\cite{Yamaguchi}  \\ 
%% 6.1 T modified  2/27/22
16.1 MeV 2$^+$T&	{\bf $5.9\times 10^{-4}$}	&	1.2	&	0.55		&	{$2.4\times 10^{-4}$}	&	2.2	&	0.6	&	- &	-	&	-	& - & -  & \\ \hline \hline
16.6 MeV 2$^-$T&	$2.6\times 10^{-4}$	&	1.6	&	0.6		&	{$5.0\times 10^{-5}$} 	&	{ 2.5}	&	{0.35}	&	- &	-	&	-	& - & -& \cite{Yamaguchi}\cite{Hicks84} \\ \hline
18.1 MeV 1$^+$T&	$1.9\times 10^{-4}$	&	0.8	&	0.35		&	$1.8\times 10^{-4}$	&	1.25	&	0.45
	&	- &	-	&	-	& - & - &\cite{Yamaguchi}\cite{Hicks84} \\ \hline\hline
{\bf 18.6 MeV-L } &	$3.2\times 10^{-4}$	&	1.3	&	0.5		&	-	&	-	&	-	&	- &	-	&	-	& - & -&\cite{Yamaguchi}  \\ \hline\hline
%%
%  OLD 19.3 MeV-T&	$7.0\times 10^{-4}$	&	1.5	&	0.65		&	$3.2\times 10^{-4}$	&	1.9	&	0.5	&	$1.0 \times 10^{-4}$ &	2.2	&	0.3	& $2.0\times 10^{-6}$ & 1 & \\ \hline\hline%%
%
{19.3 MeV 2$^-$T}&	$1.02\times 10^{-3}$	&	{ 1.32}	& { 0.77}		&	{ $3.75\times 10^{-4}$}	&	1.7	&	{0.6}	&	$1.0 \times 10^{-4}$ &	2.2	&	0.3	& $3.6\times 10^{-4}$& 1& \cite{Yamaguchi}\cite{Hicks84} \\ \hline\hline
\bf{20.0 MeV 2$^+$L} &	$1.6\times 10^{-4}$	&	1.2	&	0.42		&	$1.6\times 10^{-5}$	&	1.8	&	0.4	&	- &	-	&	-	& - & - &\cite{Yamaguchi} \\ \hline\hline
20.6 MeV 3$^-$T&	$1.9\times 10^{-4}$	&	1.45	&	0.5		&	$5.5\times 10^{-5}$	&	2.1	&	0.4	&	- &	-	&	-	& - & - & \cite{Yamaguchi}\cite{Hicks84}\\ \hline\hline
%
%\bf{(21-26)  MeV-L }&	$4.0\times 10^{-3}$	&	0.55	&	0.15		&	$6.8\times 10^{-3}$	&	0.9	&	0.63	&	- &	-	&	-	& -  & -  & &\\
%GDR $\times 0.75$	&		&	&			&		&		&		&	&		&		&  & - &\\ 
%
(21-26 MeV) &		&		&			&		&		&		&	&		&		&  & -& \\
\bf{23.0  MeV-L }&	$2.8\times 10^{-3}$	&	0.60	&	0.15		&	$6.9\times 10^{-3}$	&	0.84	&	0.55	&	- &	-	&	-	& -  & -  &\cite{Yamaguchi}\\
$\sigma$=4.75 MeV 	&		&	&			&		&		&		&	&		&		&  & -& \\ 
% \hline\hline
%(21-26)  MeV-T&	$1.7\times 10^{-3}$	&	0.7	&	0.4		&	$9.0\times 10^{-4}$	&	1.5	&	0.6	&	- &	-	&	-	& - & -&  \\
%GDR $\times 0.75$	&		&	&			&		&		&		&	&		&		&  & -& \\ 
%\hline\hline
23.0  MeV-T&	$1.83\times 10^{-3}$	&	0.8	&	0.36		&	{\bf $1.0\times 10^{-4}$}	&	1.5	&	0.5	&	- &	-	&	-	& - & - &\cite{Yamaguchi} \\
%GDR  	&		&	&			&		&		&		&	&		&		&  & - &\\ 
\hline\hline
(26-37 MeV) &		&		&			&		&		&		&	&		&		&  & -& \\
\bf{31.5 MeV-L }&	{\bf $4.7 \times 10^{-3}$}	& 1.0	&	0.48		&			& - 	&	- 	&	- &	-	&	-	& -  & - &\cite{Yamaguchi}\\
$\sigma$=9 MeV 	&		&		&			&		&		&		&	&		&		&  & - &\\ 
%\hline\hline
{31.5  MeV-T}&	{$9.0 \times 10^{-4}$}	&	{0.35}	&	{ 0.3}		&- &	-	&  - 	&	- &	-	&	-	& -  & -& \cite{Yamaguchi}\\
%GDR  &		&		&			&		&		&		&	&		&		&  & -& \\
\hline\hline
(30-50 MeV) &		&		&			&		&		&		&	&		&		&  & -& \\ 
{\bf 42  MeV-L}&	$2.6 \times 10^{-3}$	                &	1.49  & 0.7		&- &	-	&  - 	&	- &	-	&	-	& -  & - &\\
 $ \sigma$=12 MeV 	&		&		&			&		&		&		&	&		&		&  & -& \\ 
     Extra Strength &		&		&			&		&		&		&	&		&		&  & -& \\ 
\hline\hline
%\hline\hline
% \hline
 \end{tabular}
\caption{Parameterizations of the  Longitudinal (L) and Transverse (T) ${\rm ^{12}C}$  nuclear excitation form factors (squared) for the 7.65 MeV state and for states with excitation energy above 10  MeV. Unlike the  parametrizations in Table~\ref{excited_states1} for the  4.44 and 9.64 MeV states which are functions of the square of the 3-momentum transfer  ${\bf q}^2_{\rm eff}$ in units of fm$^{-2}$, the parametrizations for the states  in this table are functions of  ${\bf q}_{\rm eff}$ in units of fm$^{-1}$. Here   ${\bf q^2_{\rm eff}}={\bf q^2} \times (1+ 0.00465/E)^2$, where E is in GeV\cite{q_effective}.
}
\label{excited_states2}
\end{center}
\end{table*} 
%    TTTTTTTTTTTTTTTTTTTTTTTTTTTTTTTTTTTTTTTTTTTTTTTTTTTTTTTTTTT
%
% Fig. 6    FFFFFFFFFFFFFFFFFFFFFFFFFFFFFFFFFFFFFFFFFFF
\begin{figure*}
%[ht]
\begin{center}
\includegraphics[width=3.0 in,height=2.4in]{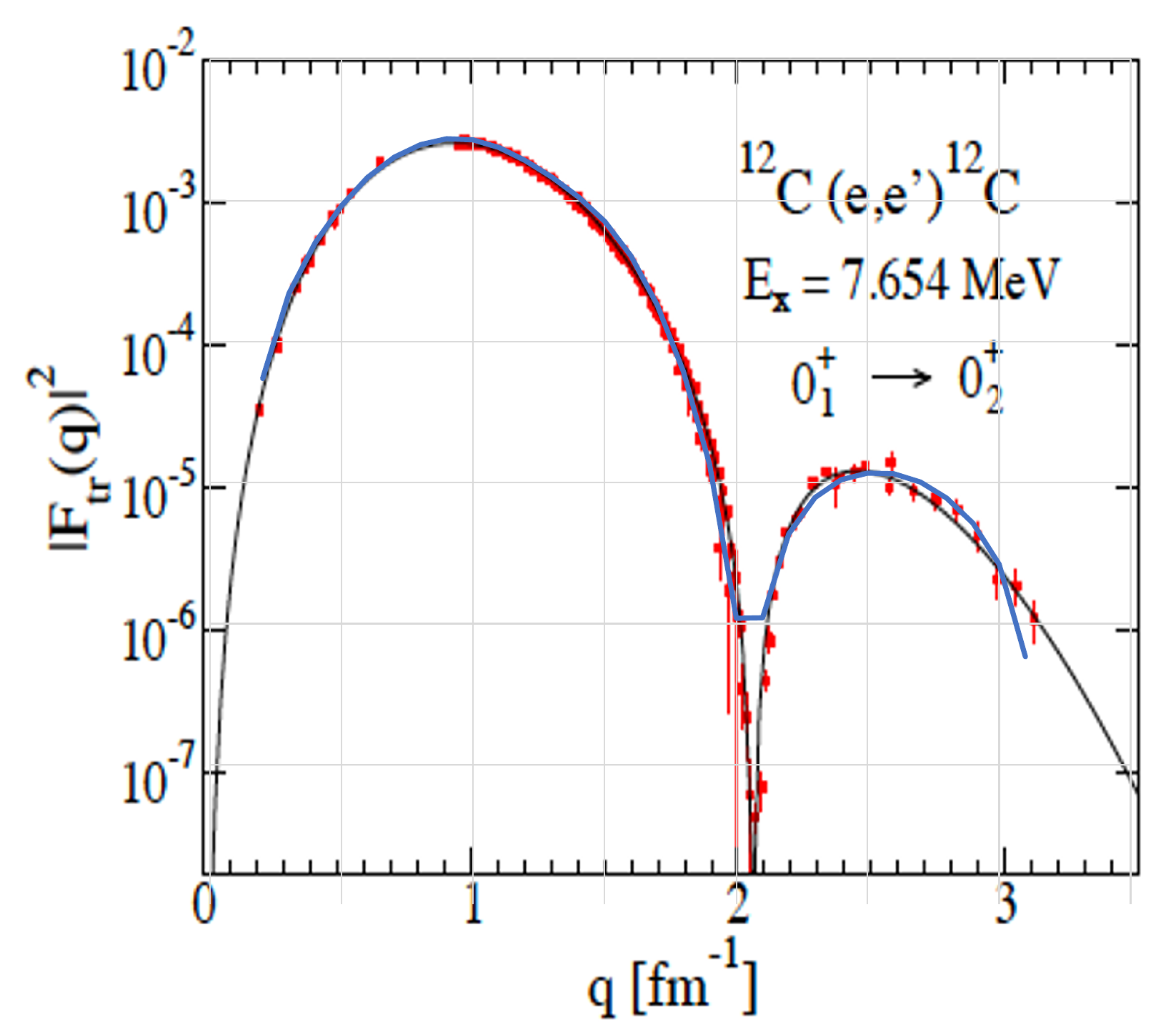}
\includegraphics[width=3.0 in,height=2.4in]{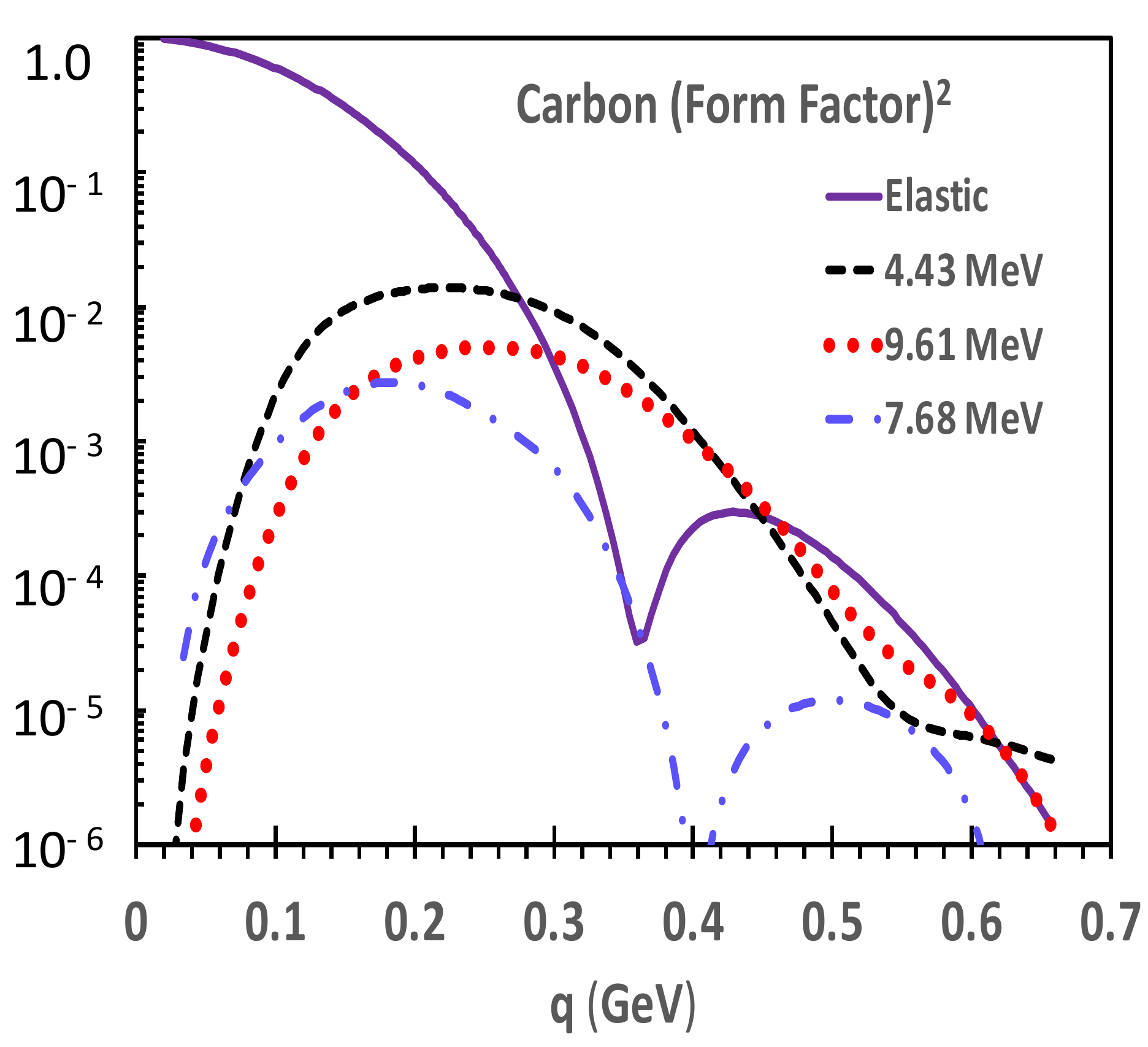}
\caption{{\bf Left panel}:  Measurements\cite{FF_765} of the longitudinal charge form factor (squared)  for the 7.65 MeV state in  ${\rm ^{12}C}$. {\bf Right panel}:  A comparison of the nuclear elastic form factor (squared) to the form factors (squared) of the first three nuclear excitations versus $\bf q$ (in units of GeV). }
\label{C12states1}
\end{center}
\end{figure*} 
%
%
%FFFFFFFFFFFFFFFFFFFFFFFFFFFF
% Fig. 7
\begin{figure*}
\begin{center}
\includegraphics[width=3.5in,height=5. in]{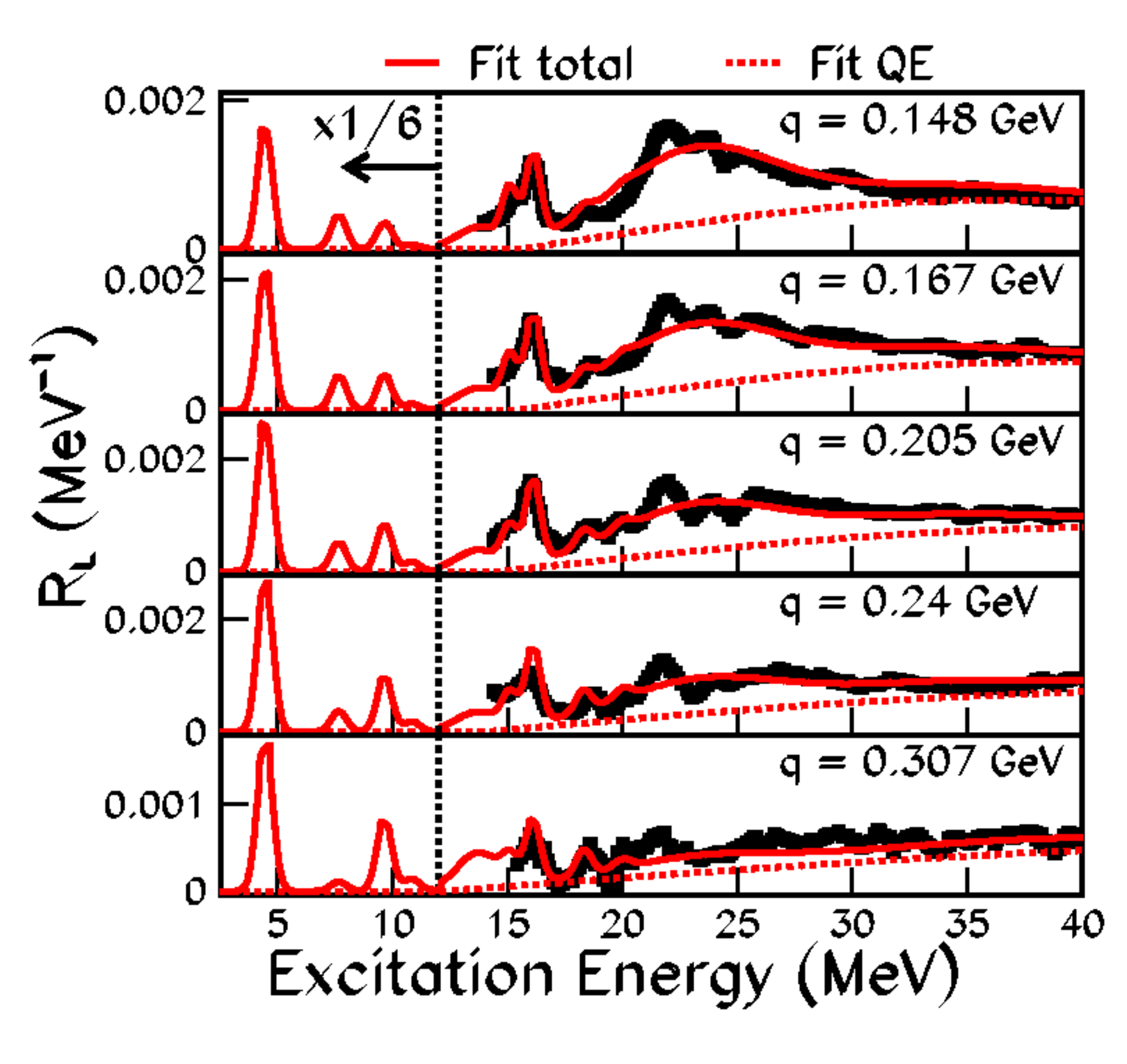}
\includegraphics[width=3.5in,height=5. in]{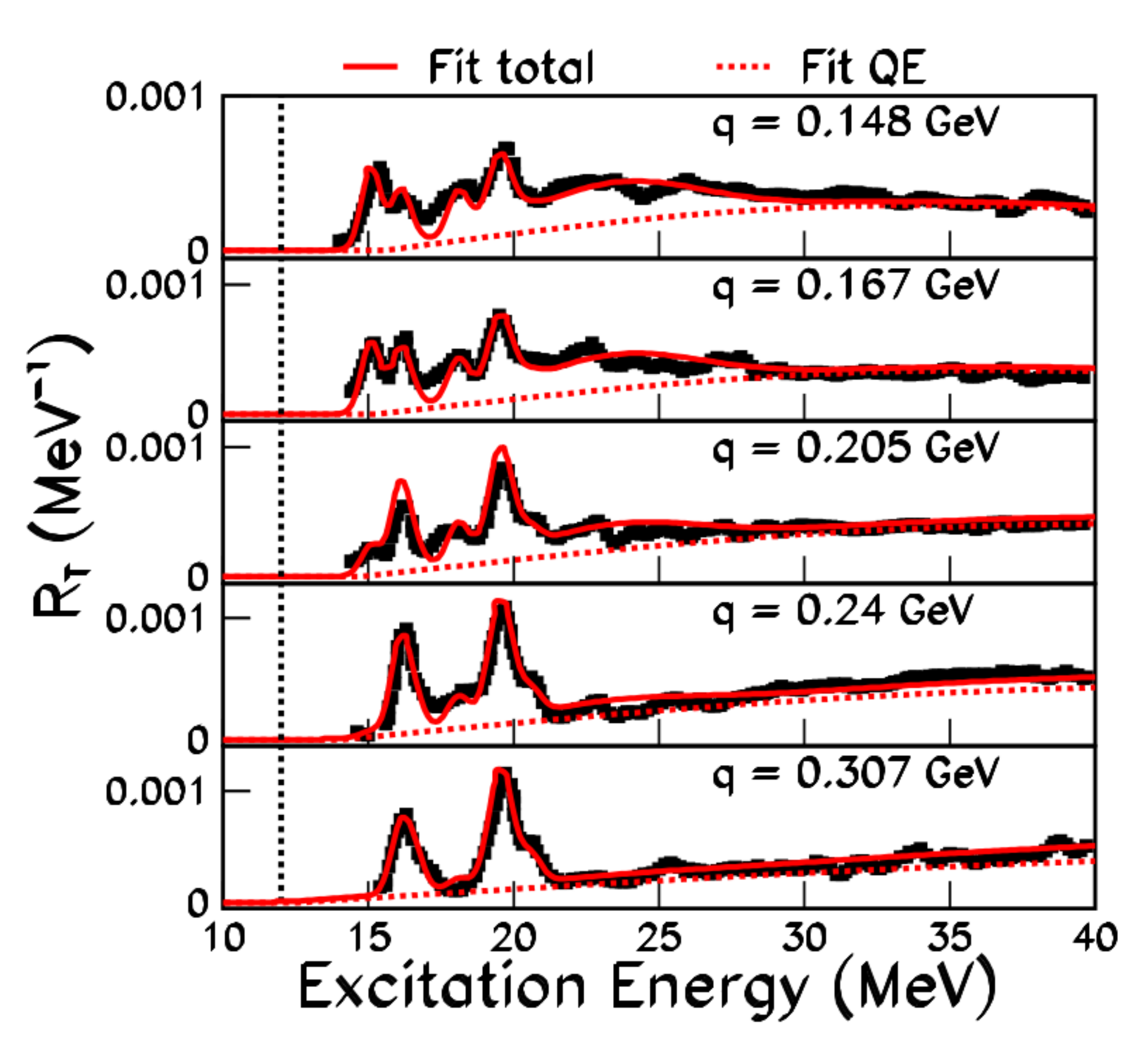}
\caption{Comparison of the longitudinal ($R_L$, left) and transverse ($R_T$, right) response functions for ${\rm ^{12}C}$ extracted by Yamaguchi 71\cite{Yamaguchi} (black squares) to the response functions extracted from our  universal fit to all available electron scattering cross section data on ${\rm ^{12}C}$ (solid red line).
The  contributions from excitation energies less than 12 MeV are multiplied by (1/6).  The QE contribution to the total response functions is represented by the red dashed line.   In our fit,  we  model  the response functions for all states the  region of the Giant Dipole Resonance (20-30 MeV) region as one average broad excitation.}
\label{Yamaguchi_RL_RT}
\end{center}
\end{figure*} 
%FFFFFFFFFFFFFFFFFFFFFFFFFFFF
%TTTTTTTTTTTTTTTTTTTTTTTTTTTTTTTTTTTTTTTTTTTTTTTTTTT
%FFFFFFFFFFFFFFFFFFFFFFFFFFFF
% Fig. 8
\begin{figure*}
\begin{center}
\includegraphics[width=7.0in,height=8.0in]{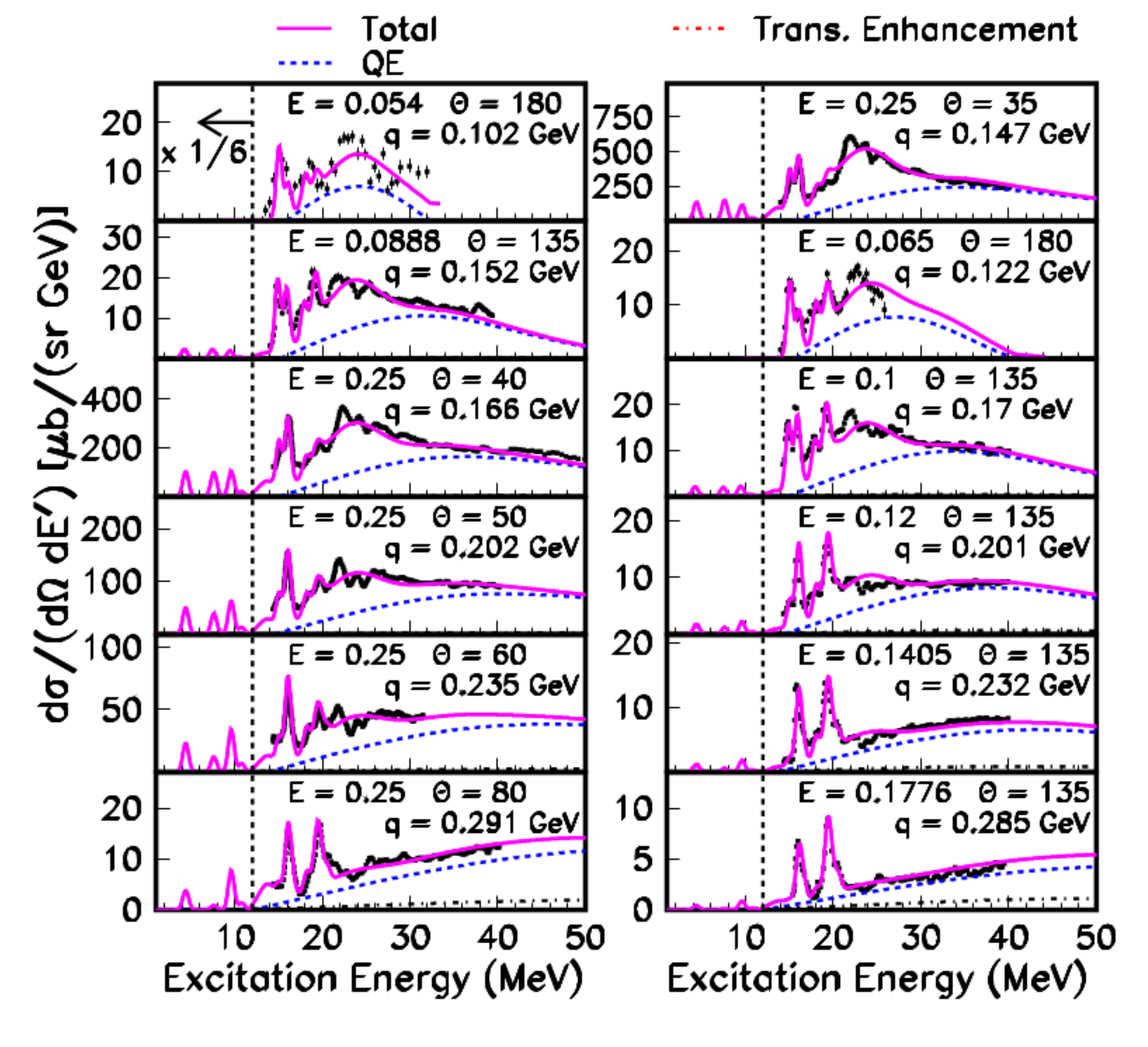}
\caption{Radiatively corrected inelastic electron scattering cross sections on ${\rm ^{12}C}$ for excitation energies less than 50 MeV.  The  cross sections for excitation energies less than 12 MeV are multiplied by (1/6). The pink solid line is the predicted total cross section from our  universal fit\cite{short_letter} to all electron scattering data on ${\rm ^{12}C}$.  The fit include nuclear excitations, a superscaling QE model\cite{superscaling} with Rosenfelder Pauli suppression\cite{Rosenfelder}  (dashed blue line), "Transverse Enhancement/Meson Exchange Currents" (dot-dashed line) and pion production processes (at higher excitation energies). 
%Details of the fit are described in reference \cite{short_letter} 
The data are from Yamaguchi71\cite{Yamaguchi} except for  the cross sections  for $E_o$=54 MeV and 180$^0$ (from Goldemberg64\cite{Goldemberg64}) and the cross sections for  for $E_o$=65 MeV and  180$^0$ (from deForest65\cite{deForest65}). The measurements at 180$^0$ are only sensitive to the transverse form factors.}
\label{Yamaguchi_states}
\end{center}
\end{figure*} 
%FFFFFFFFFFFFFFFFFFFFFFFFFFFF
%  SBSBSBSBSB sub Section A.2
\subsection {Form factors for nuclear excitations in ${\rm ^{12}C}$} 
We begin by parameterizing the measurements of the  longitudinal and transverse  form factors for the electro-excitation of all  nuclear states in ${\rm ^{12}C}$ with excitation energies ($E_x$) less than 16.0 MeV (the approximate proton removal energy  from ${\rm ^{12}C}$). For these states the measurements are straightforward since the QE cross section is zero for $E_x<$ 16 MeV.  
    \subsubsection {${\rm ^{12}C}$ excitation form factors for the 4.44 MeV and 9.64 MeV states} 
  The  longitudinal form factors (squared) for the electro excitation of the 4.44 and 9.64 MeV nuclear  excited states are parametrized as $F_{iC}^2({\bf q}^2_{\rm eff})$ where
  % eq. 25
   \begin{equation}
 \label{verses_q2}
F_{iC}^2({\bf q}^2_{\rm eff})= \frac{({\bf q^2}_{\rm eff})^3}{({\bf q^2}_{\rm eff})^3+d } \sum_{j=1}^{j=3}  N_j e^{-[({\bf q}^2_{\rm eff}-C_j)/\sigma]^2}
\end{equation}
%% \begin{equation}
% \label{verses_q2}
%F_i^2({\bf q^2})=  \sum_{j=1}^{j=3}  N_j e^{-[({\bf q}^2-C_j)/\sigma]^2} - a e^{-b{\bf q}^2}.
%\end{equation}V
Here ${\bf q}^2$ is  in units of fm$^{-2}$.
The parameters for the  4.44 and 9.64 MeV states are given in Table \ref{excited_states1}. Comparisons of our  parametrizations of  the excitation form factors (squared) for the 4.44, and 9.64 MeV  states  to experimental data\cite{C12_FF} are shown in Figures  \ref{C12states} and  \ref{C12_ratios}.
%
%
%-------------------------------------------------------------------------------
%
  \subsubsection  {${\rm ^{12}C}$ form factors for the 7.65 MeV state and states with excitation energies above 10 MeV}
   Measurements of the square of the longitudinal form factor verses $\bf q$ (in units of fm$^{-1}$) for the  7.65 MeV state  in ${\rm ^{12}C}$  (from Chernykh et. al. \cite{FF_765}) are shown on the left  panel of Fig.  \ref{C12states1}.  
A comparison of the nuclear elastic form factor to the form factors of the first three nuclear excitations  versus $\bf q$ (in units of GeV) is shown on the right panel of Fig.  \ref{C12states1}.

 The charge  form factors  (squared) for the electro-excitation of  the  7.65 MeV  state and for  states with  excitation energies above 10 MeV are  parameterized  as  $F_{iC}^2({\bf q})=Max(0.0,g_i^2)$ where
 %
 %Eq 26
  \begin{equation}
  \label{verses_q}
g_i^2({\bf q_{eff}})=   \sum_{j=1}^{j=3}  N_j e^{-[({\bf q_{eff}}-C_j)/\sigma]^2} - a e^{-b{\bf q_{eff}}}.
\end{equation}
Here,  $\bf q_{eff}$ is in units of fm$^{-1}$.  The parameters are given in Table \ref{excited_states2}. 
(Note that these states are parameterized versus  $\bf q_{eff}$, while the 4.44 and 9.64 MeV states are parametrized versus ${\bf q_{eff}}^2$). 
As shown on the right panel of Fig. \ref{C12states1}, for $\bf q$ near the  diffraction minimum for elastic scattering on ${\rm ^{12}C}$  the cross sections for the three nuclear excitations below 10 MeV are larger than the nuclear elastic cross section. 
 %Because of initial state photon emission, the  modeling of the cross sections for the excitation of nuclear states is needed for calculations of radiative corrections for QE scattering at low  $\bf q$, and  deep-inelastic scattering at large values of $\nu$.
 Note that unlike the nuclear elastic form factor which is equal to 1.0 at $\bf q$=0, all longitudinal form factors for the nuclear excitations must vanish at $\bf q$=0.
\subsubsection  {${\rm ^{12}C}$ form factors for states with excitation energies above 10 MeV and below 16 MeV}
We use equation \ref{verses_q}  to parameterize the form factor for excitation energy of  
10.84 MeV\cite{FF_1084}, and also  for excitation energies of  12.71, 14.09 and 15.11 MeV\cite{FF_1271_1511, FF_1408}.  In addition, we find that published differential cross section measurements indicate that there is an additional longitudinal continuum in the region between  12 to 15 MeV. We parameterize this longitudinal continuum as one broad state at 13.7 MeV ($\sigma$=1.25 MeV). For the transverse form factors in this region we parametrize the data of Hicks84\cite{Hicks84}.
\subsubsection  {${\rm ^{12}C}$ form factors for states with excitation energies above 16 MeV}
  Initially, we parameterize the  longitudinal and transverse form factors  measured by Yamaguchi71\cite{Yamaguchi} for  states with excitation energies above 16 MeV. However, in the Yamaguchi71 analysis the  contributions from quasielatic (QE) scattering are not subtracted. Therefore, We perform a reanalysis of the Yamaguchi71 data in combination of all published cross sections with 16~$<E_x<$~55 MeV. We subtract the QE contribution using our QE model\cite{short_letter} (which includes superscaling\cite{superscaling} with Rosenfelder\cite{Rosenfelder} Pauli Suppression) 
  and extract updated longitudinal and transverse form factors. For $E_x>$~20 MeV (region of the Giant Dipole resonances) we group the strength from multiple excitations into a three states with a large width in $E_x$ and extract effective form factors accounting for all states in these regions.  
  The updated parameters are given in Table \ref{excited_states2}.  
  
  The   longitudinal and transverse response functions for ${\rm ^{12}C}$, $R_L({\bf q},E_x)$ and $R_T({\bf q},E_x)$, extracted by Yamaguchi71\cite{Yamaguchi} for excitation energies above 16 MeV and less than 40 MeV are shown in Figure \ref{Yamaguchi_RL_RT} (black points). 
  Also shown are  $R_L({\bf q},E_x)$ and $R_T({\bf q},E_x)$  
  %the corresponding  longitudinal (RL) and transverse (RT) response functions for ${\rm ^{12}C}$ 
  extracted from our universal fit to all electron scattering cross section data on  ${\rm ^{12}C}$ (solid red line). The QE contribution to the total response functions is shown as the dashed red line. An estimated resolution smearing of 600 keV has been applied to the excitations in the fit to match the data. While individual states are well reproduced at low excitation energy, above $E_x$ of 20~MeV the effect of  grouping several excitations together into three broad effective states in the fit can be seen.  While the fit does not capture the structure from individual states above 20 MeV, the total strength is seen to be well reproduced.
  \subsection {Comparison to ${\rm ^{12}C}$ experimental data for excitation energies below 50 MeV}
%
%Comparison of our fits to some of the  experimental data with excitation energies below 20 MeV are  shown in Figures \ref{LEDEX_fig},  \ref{Yamaguchi_states} and \ref{Yamaguchi_RL_RT}

 Experimental radiatively corrected inelastic electron scattering cross sections on ${\rm ^{12}C}$ for excitation energies less than 50 MeV are shown in Figure \ref{Yamaguchi_states}. Also shown are the corresponding cross sections from our universal fit to all ${\rm ^{12}C}$ data.  The  cross sections for excitation energies less than 12 MeV are multiplied by (1/6). The pink solid line is the predicted total cross section from our universal  fit\cite{short_letter} which include  the contributions from all sources (nuclear excitation form factors, quasilelastic scattering and pion production processes).  The QE contribution is shown as the dashed blue line and the  "Transverse Enhancement/Meson Exchange Currents" contribution is shown as the dot-dashed  line. Details of the fit are described in reference\cite{short_letter}.  Most of  the cross section measurements  are from Yamaguchi71\cite{Yamaguchi}. The cross sections for $E_o$=54 MeV at 180$^0$ are from Goldemberg64\cite{Goldemberg64} and the the cross sections for $E_o$=65 MeV at 180$^0$ are from deForest65\cite{deForest65}.  The measurements at 180$^0$ are only sensitive to the transverse form factors.                   
%
%
%
%SSSSSSSS SECTIOM
%
%Fig. 9
\begin{figure*}
%\vspace{9pt}
\begin{center}
\includegraphics[width=2.3in, height=2.95 in]{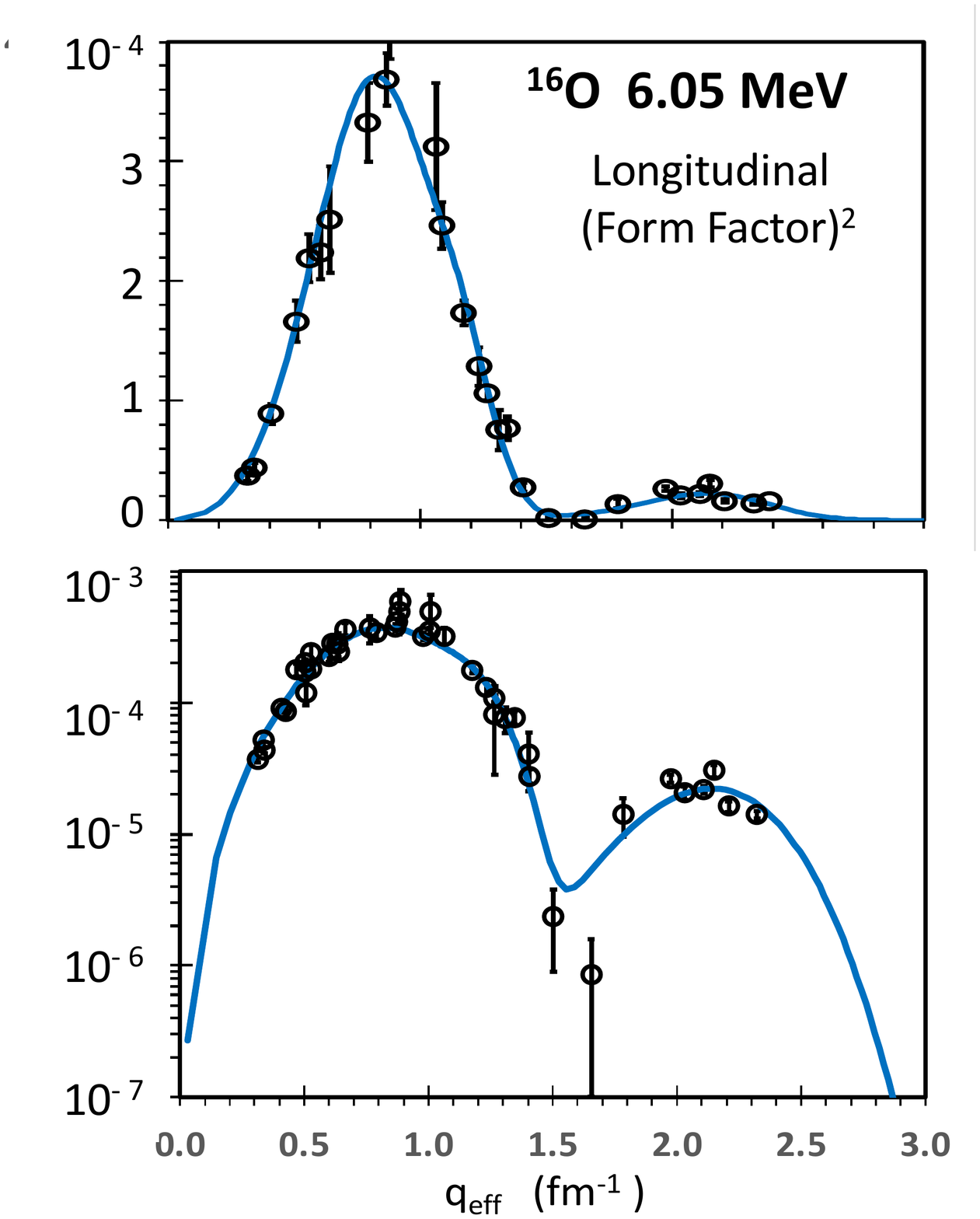}
\includegraphics[width=2.3in, height=2.95 in]{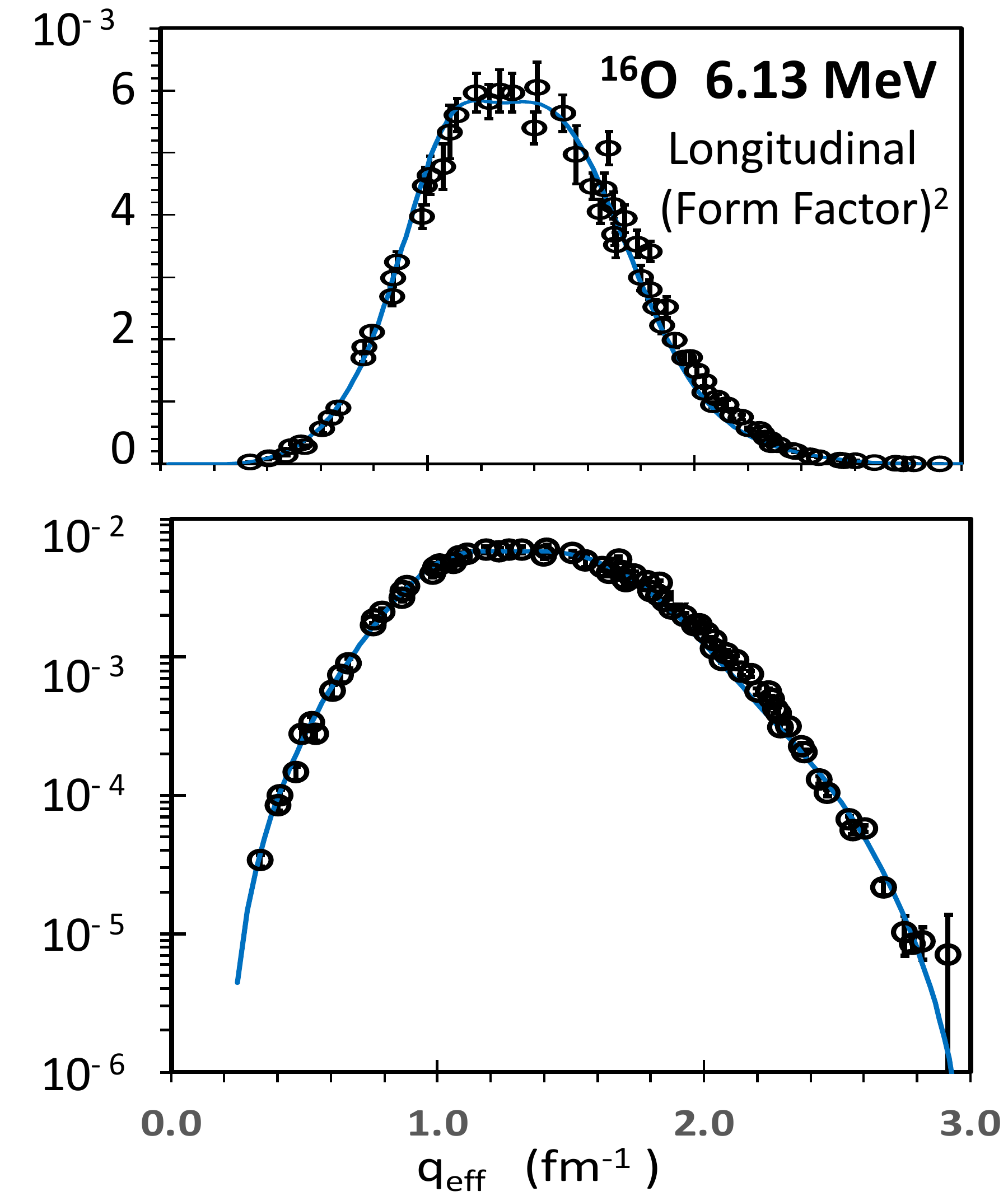}
\includegraphics[width=2.3in, height=2.95 in]{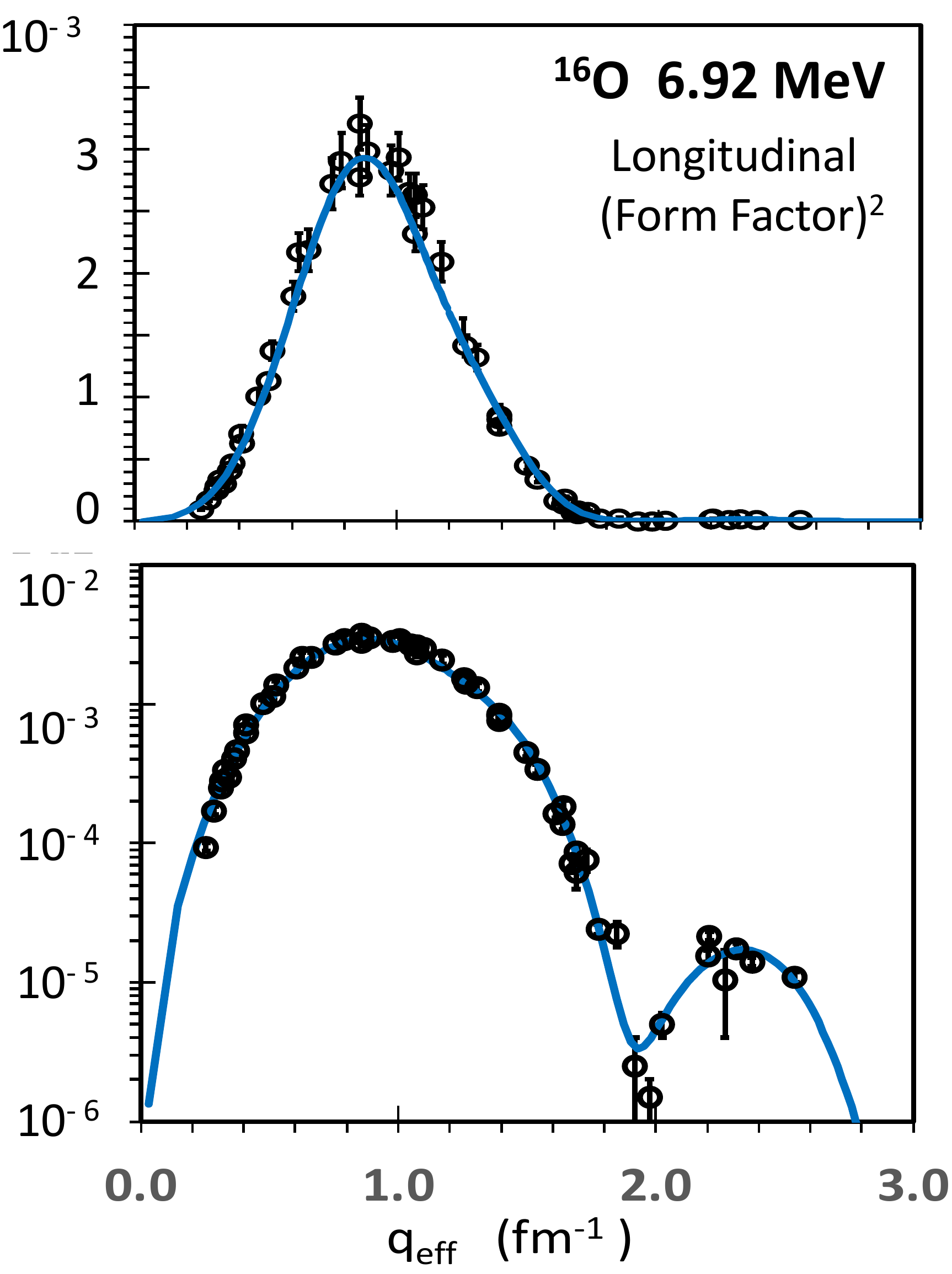}
\includegraphics[width=2.3in, height=2.95 in]{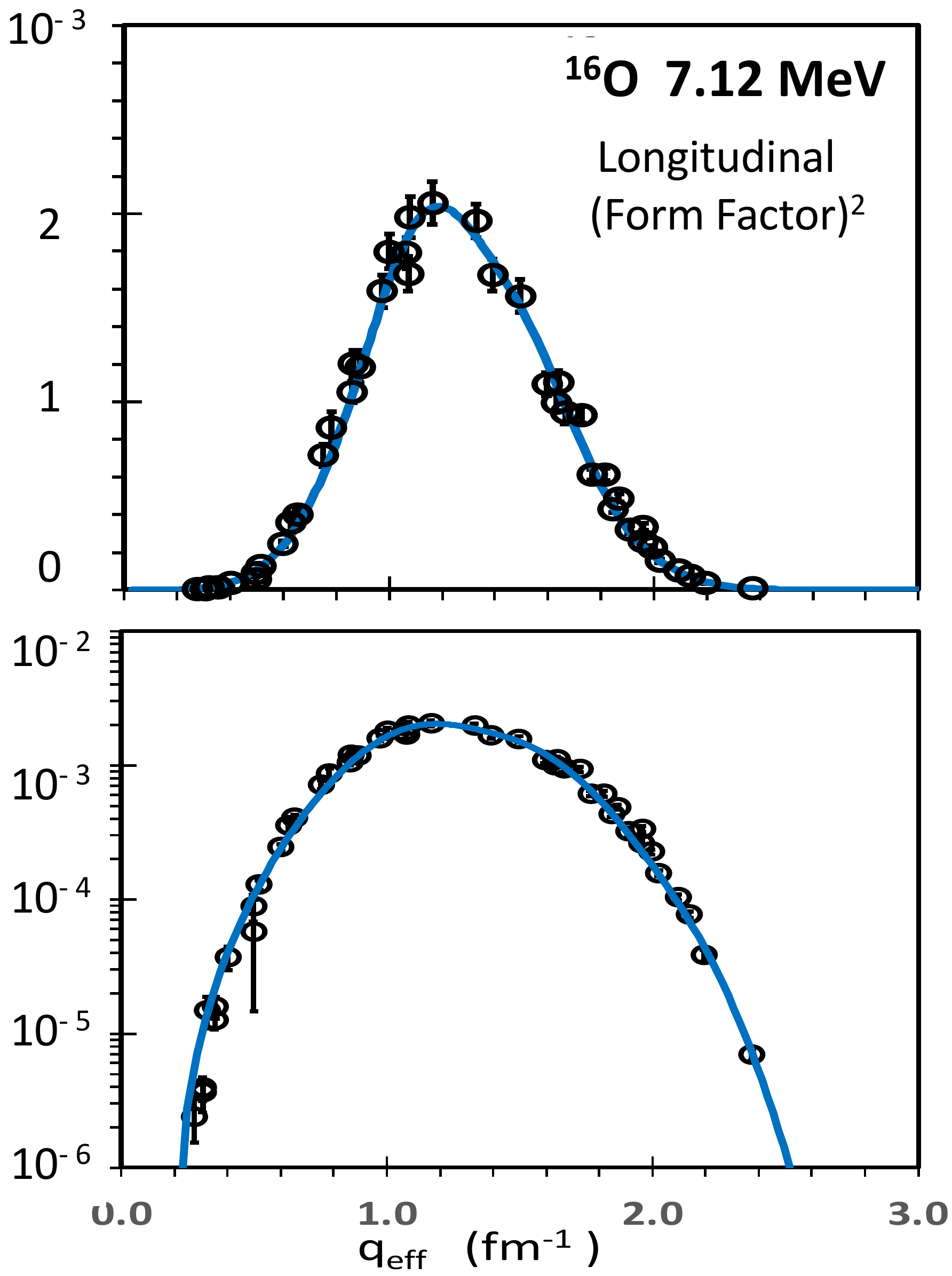}
\includegraphics[width=2.3in, height=2.95 in]{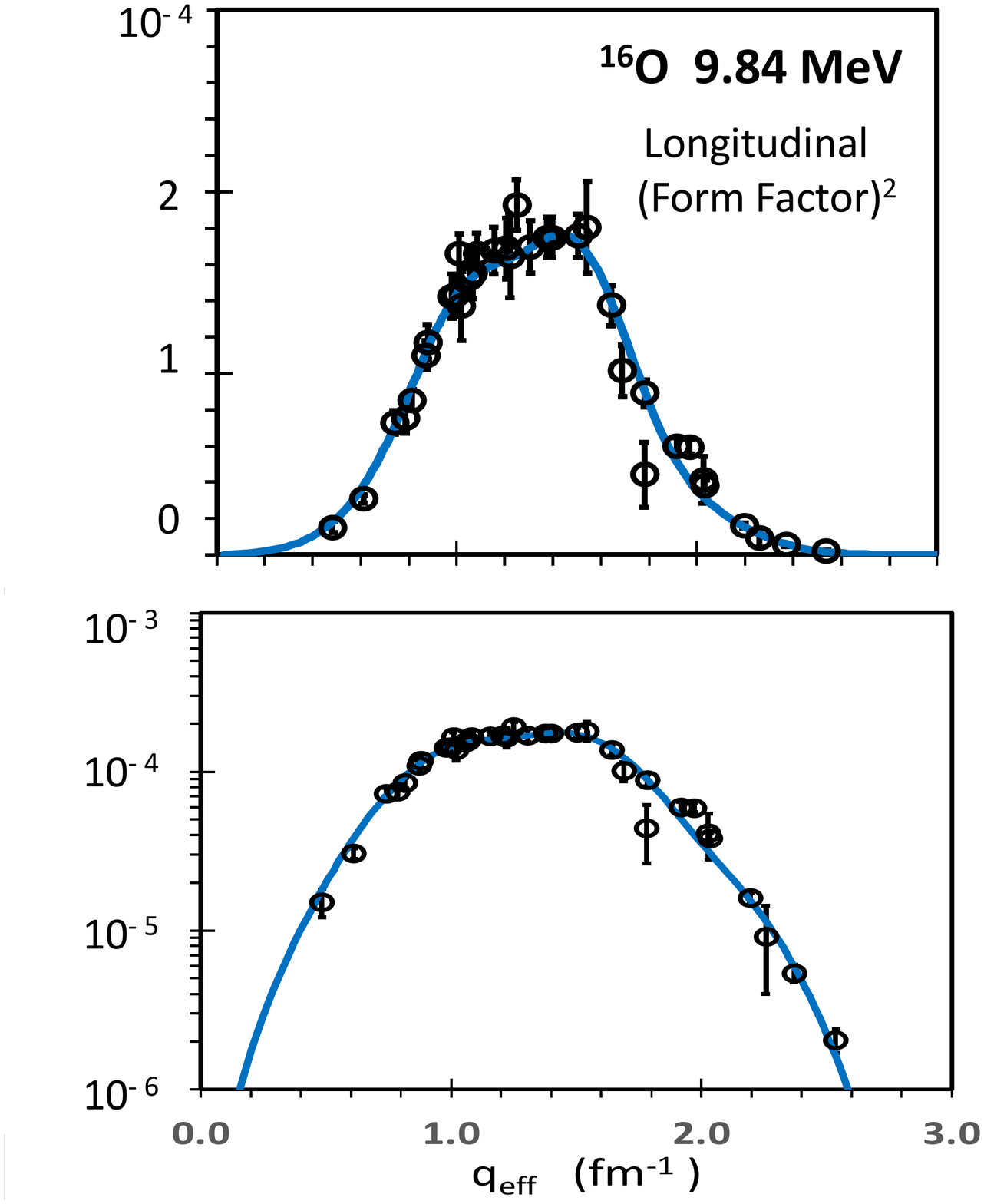}
\includegraphics[width=2.3in, height=2.95 in]{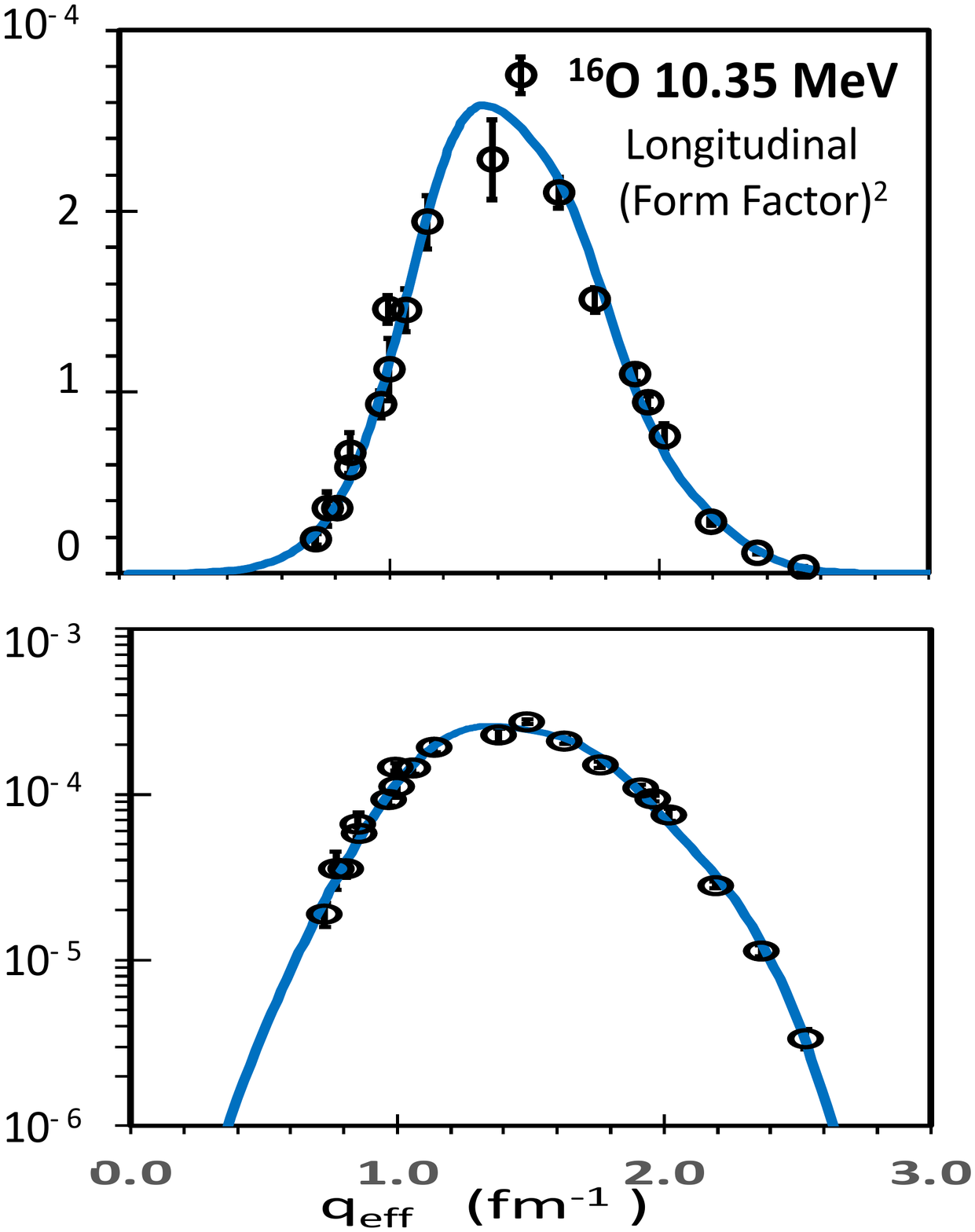}
\includegraphics[width=2.3in, height=2.95 in]{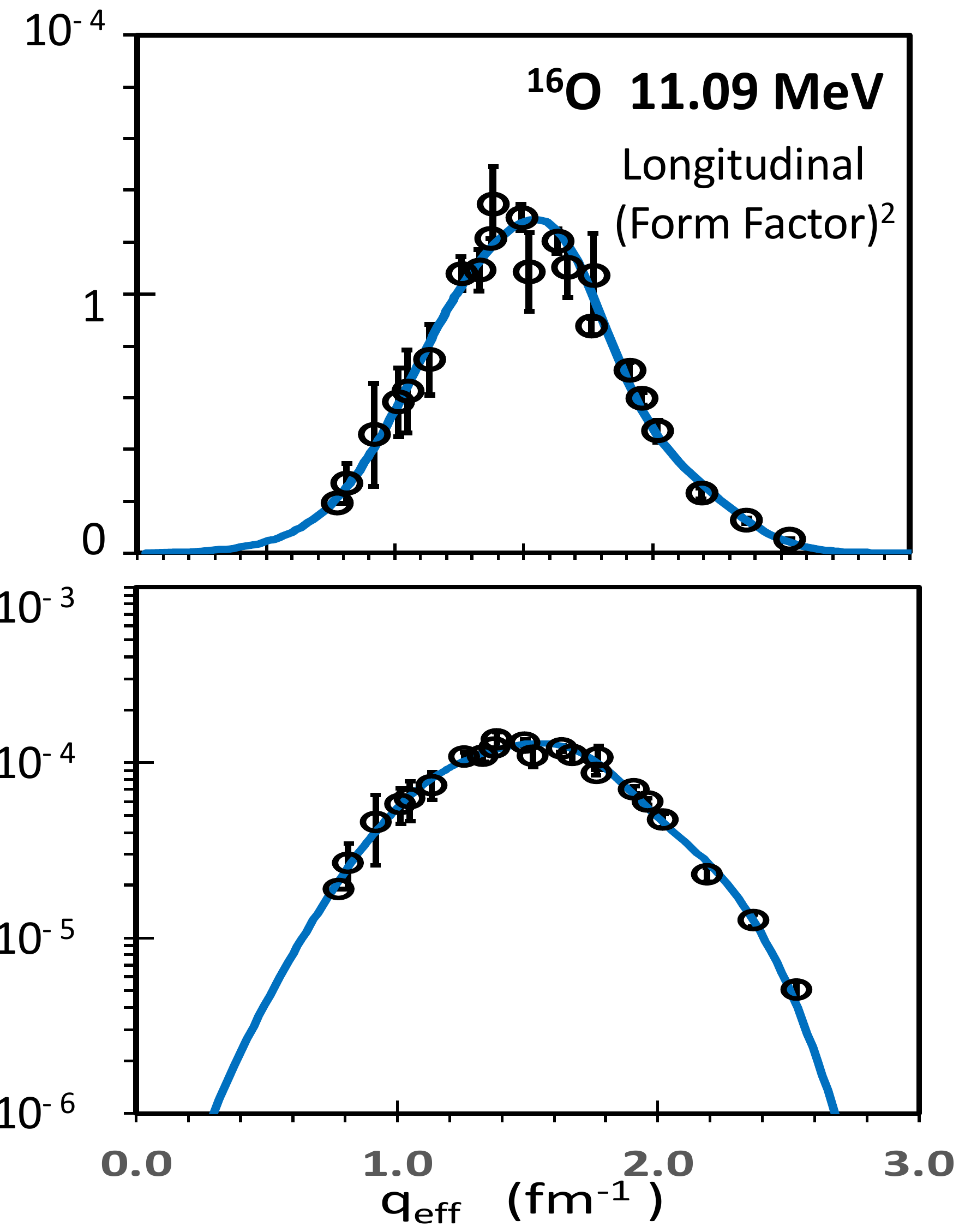}
\includegraphics[width=2.3in, height=2.95 in]{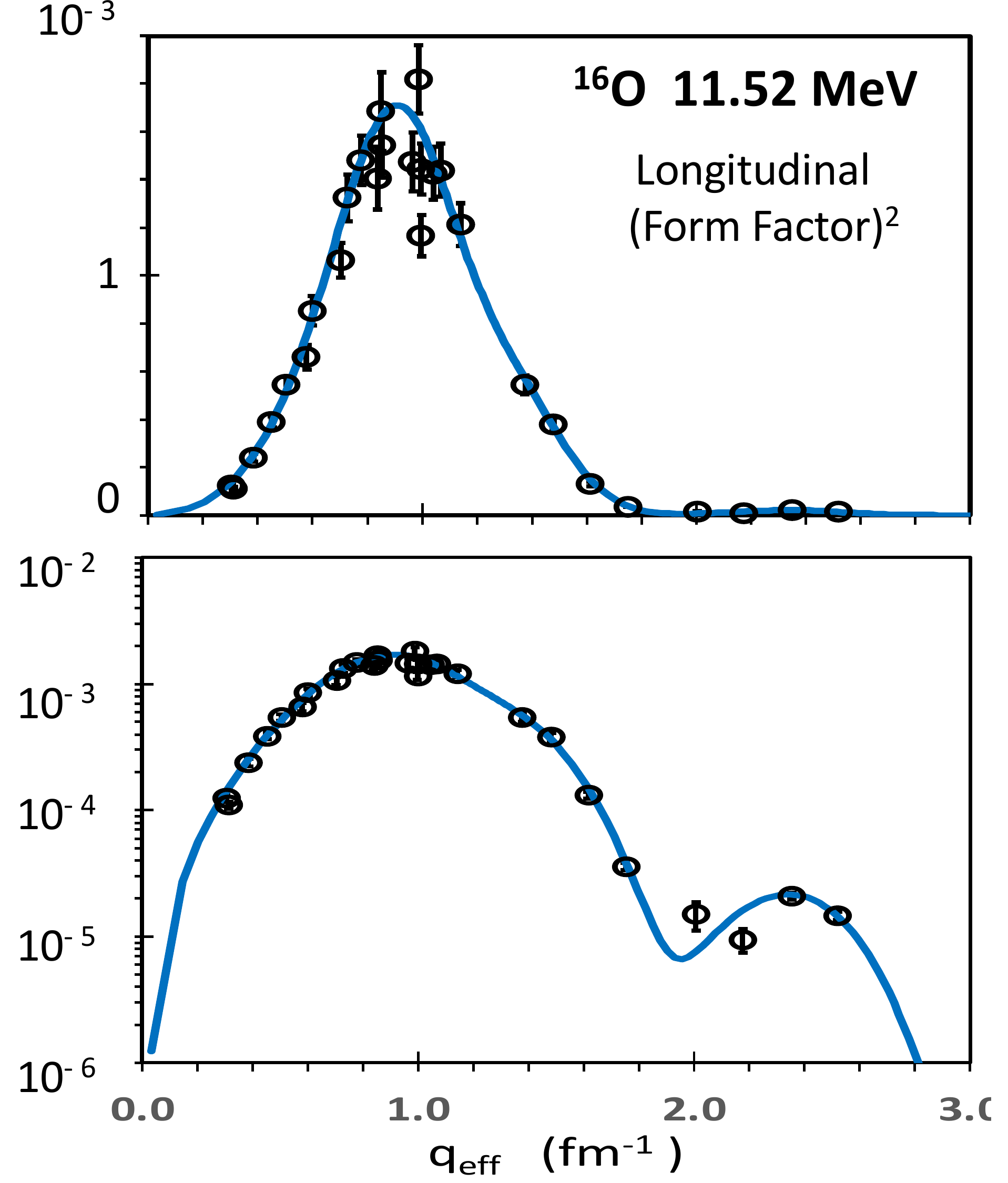}
\includegraphics[width=2.3in, height=2.95 in]{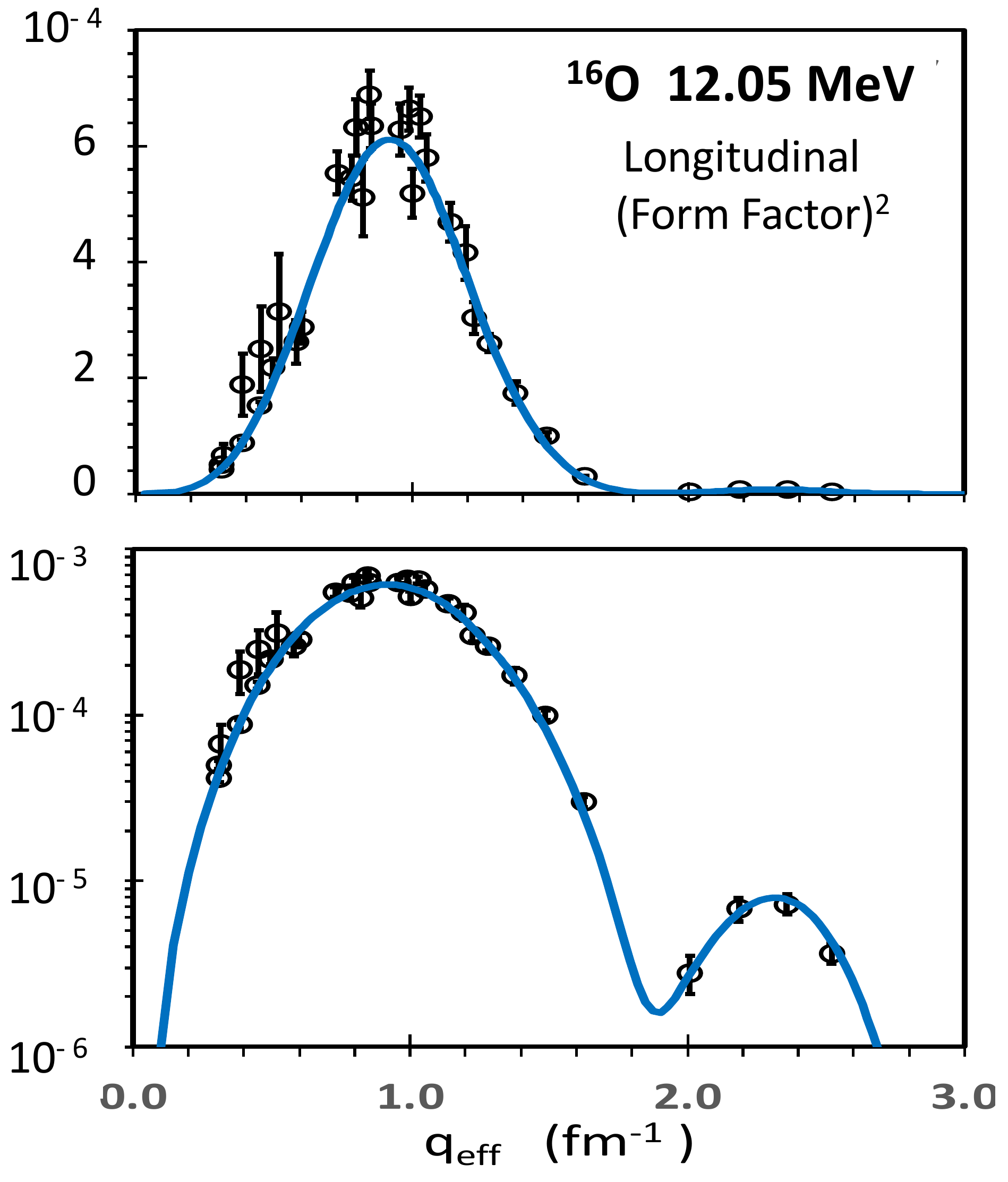}
\caption { The square of the longitudinal form factors  in  ${\rm ^{16}O}$  for nuclear excitations below 12.5 MeV on linear and logarithmic scales. The data are from Buti-86\cite{Buti86}. The  blue solid lines are our parameterizations from Table \ref{O16_states}.}
\label{O16_Buti_states}
\end{center}
\end{figure*}

%% Fig. 10  FFFFFFFFFFFFFFFFFFFFFFFFFFFFFFFFFFFFFFFFFFF
%
\begin{figure*}
%\vspace{9pt}
\begin{center}
\includegraphics[width=2.3 in, height=2.5 in]{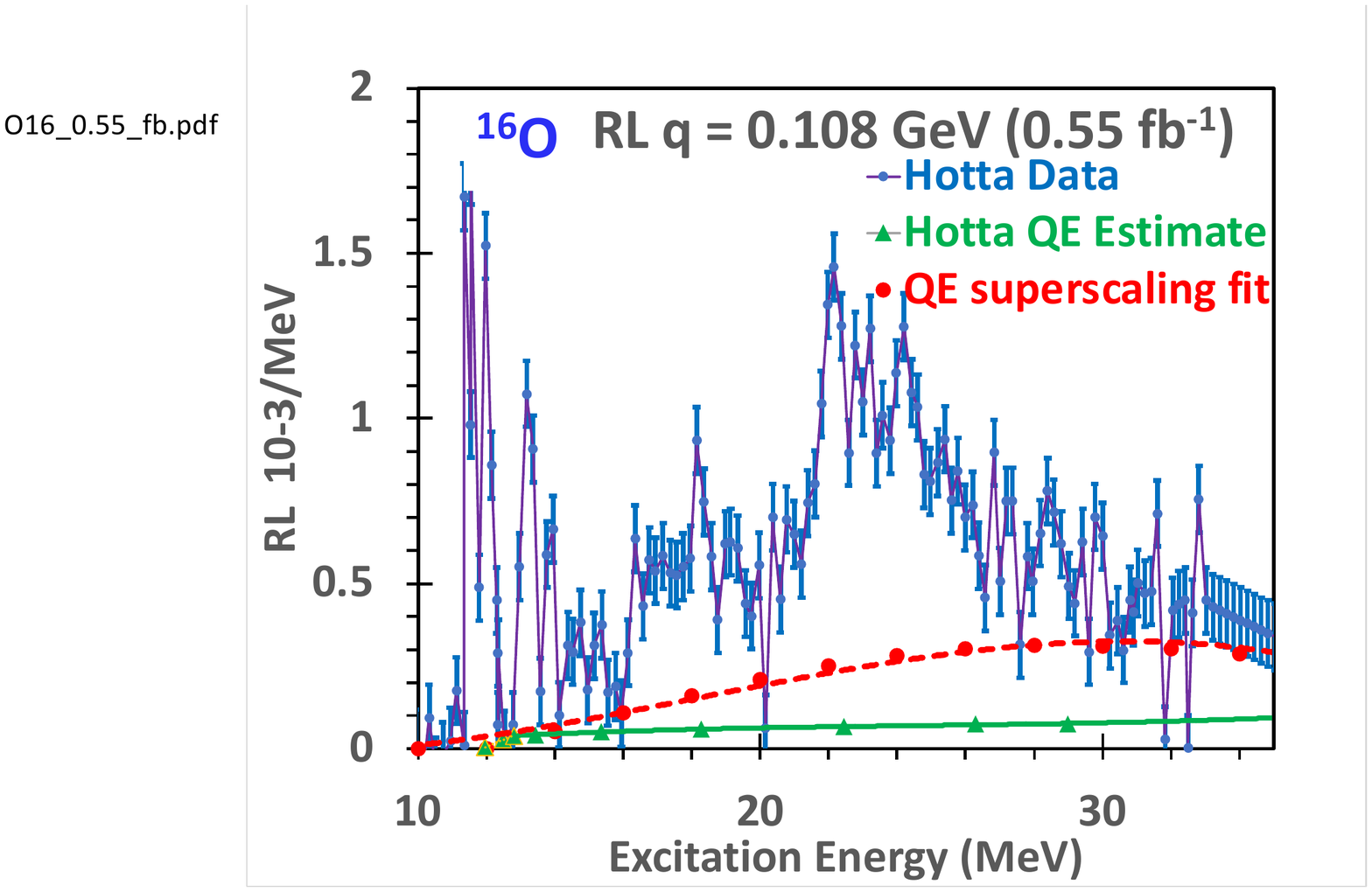}
\includegraphics[width=2.3 in, height=2.5 in]{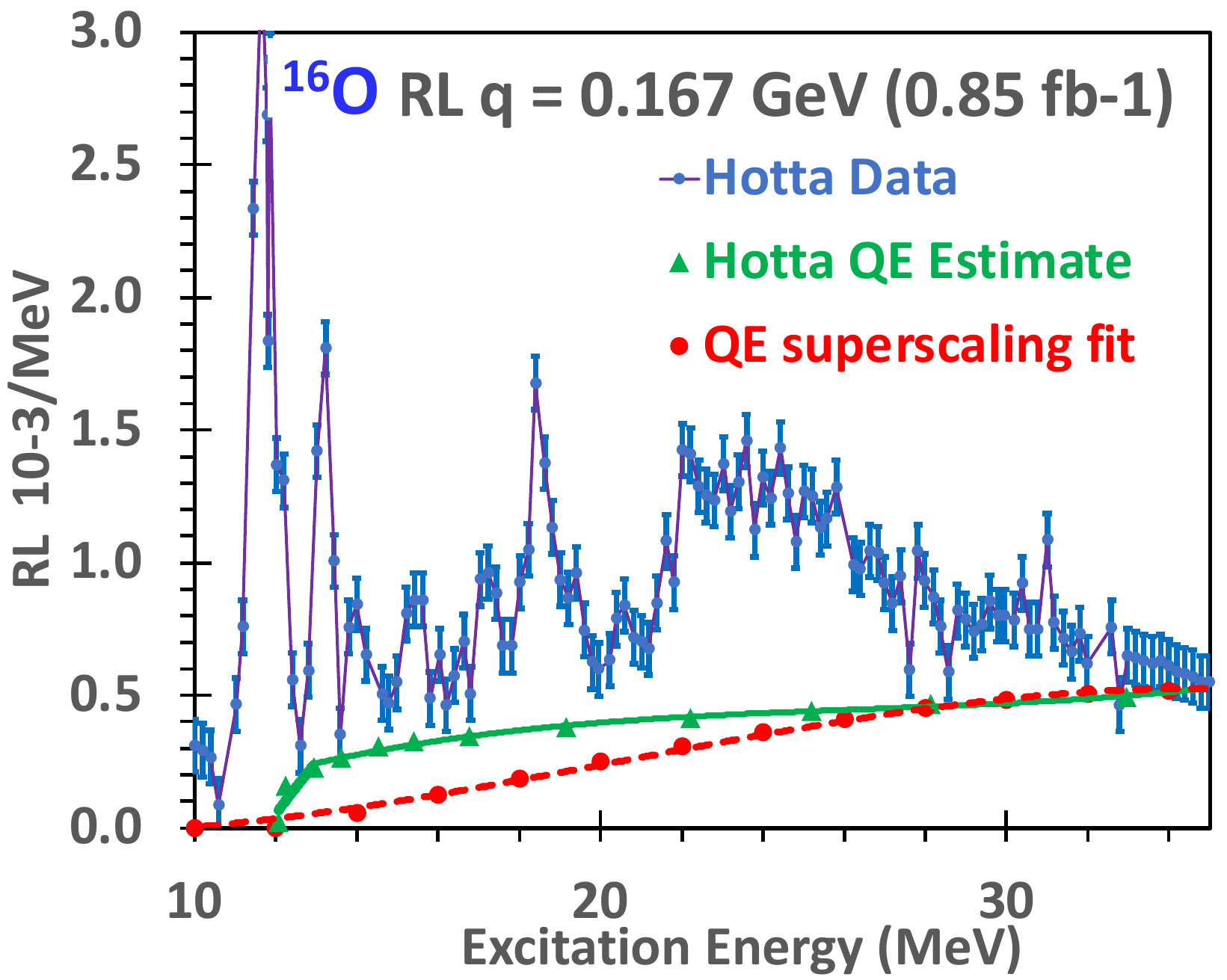}
\includegraphics[width=2.3 in, height=2.5 in]{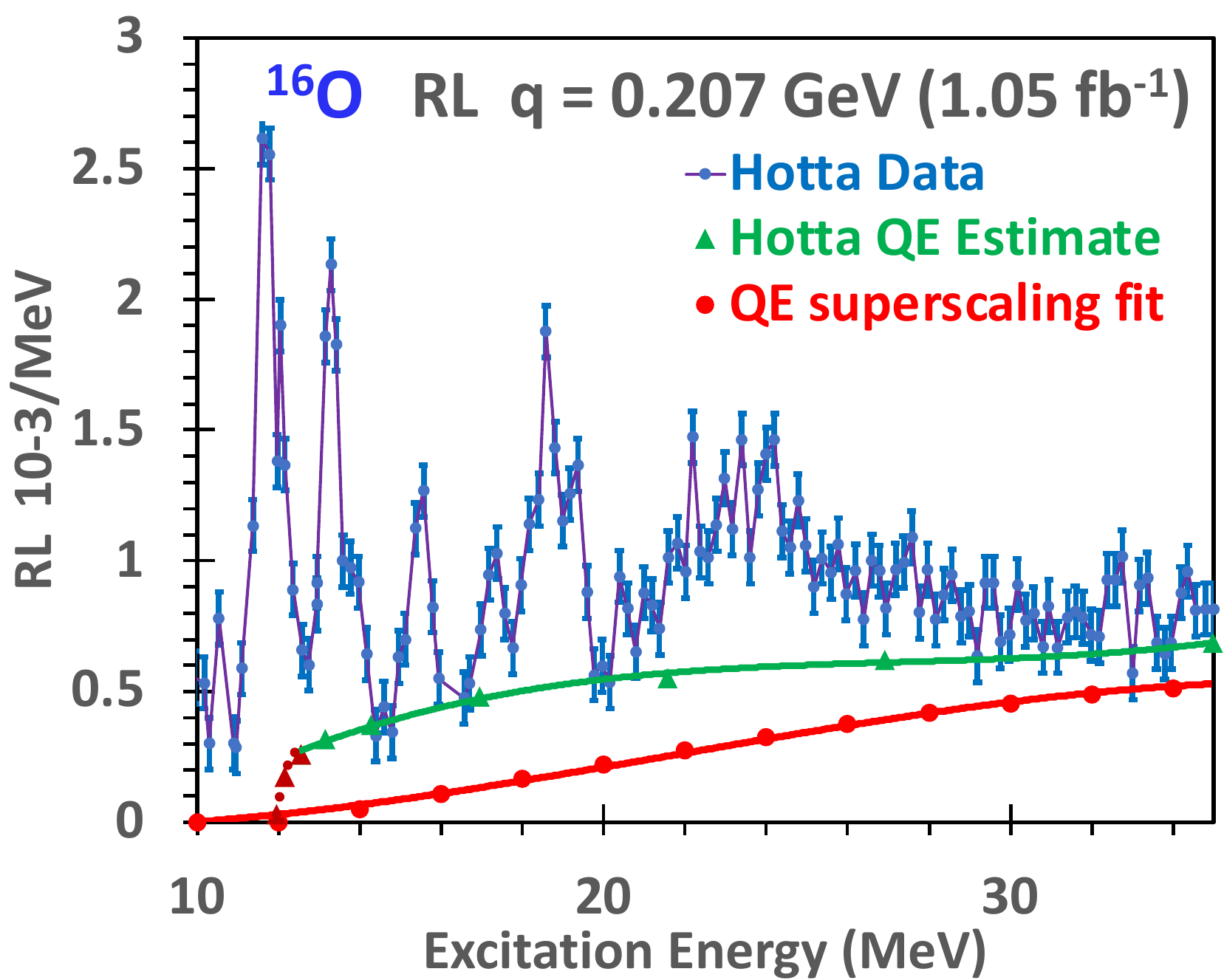}
\includegraphics[width=2.3 in, height=2.95 in]{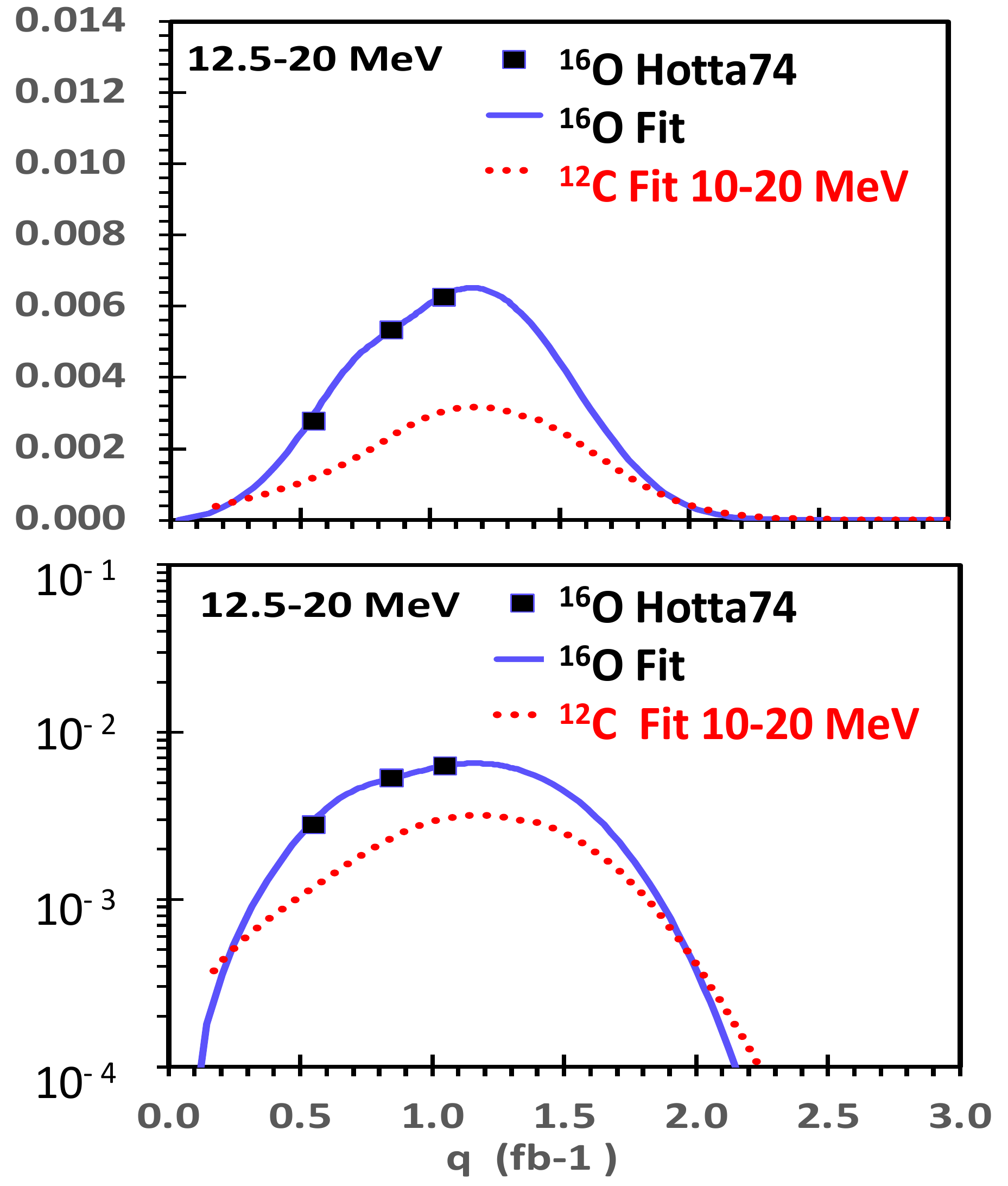}
\includegraphics[width=2.3 in, height=2.95 in]{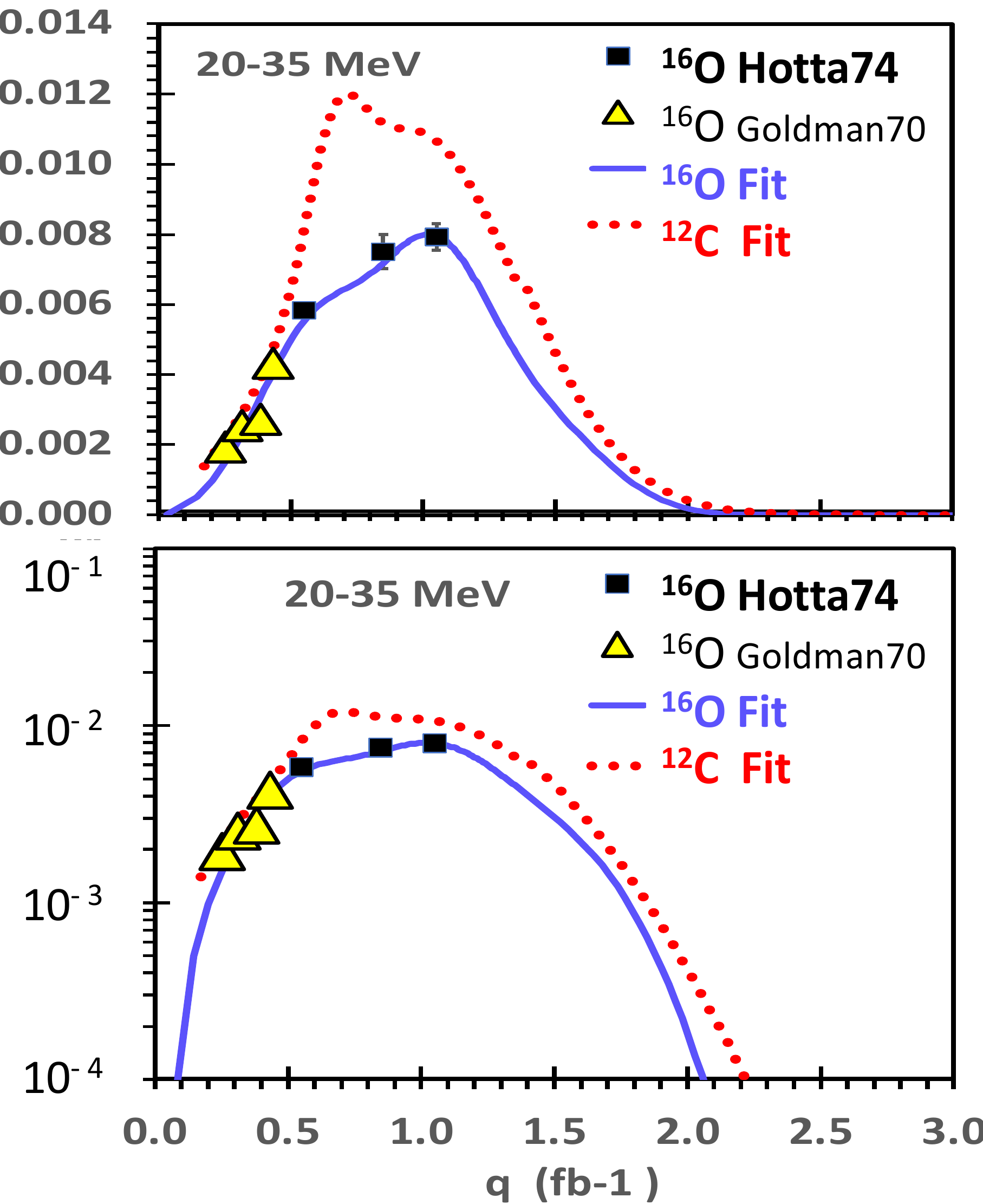}
\includegraphics[width=2.3 in, height=2.95 in]{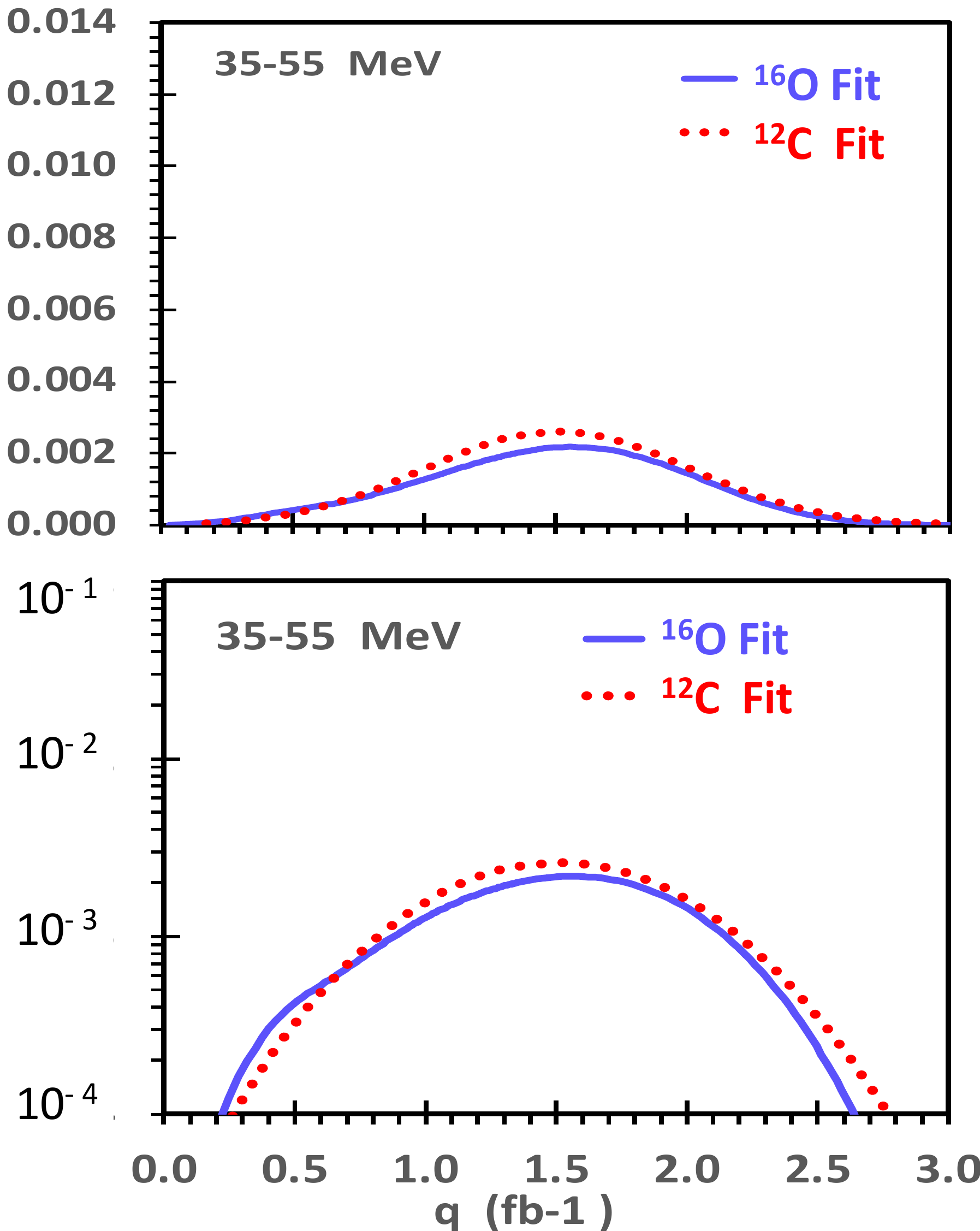}
\caption{
{\bf Top row}: The longitudinal response function ${\rm ^{12}C}$, $R_L({\bf q},E_x)$ for ${\rm ^{16}O}$  from Hotta74\cite{Hotta74} for three values of $\bf q$. The red dashed line  is the original estimate of the QE contribution used in  Hotta74.  The green solid line is the QE contribution determined using our superscaling model.
We use these data to extract the longitudinal form factors for  nuclear excitations in ${\rm ^{16}O}$ for the 12.5-20 MeV and the 20-35 MeV regions in excitation energy.
{\bf Middle and bottom rows}: The $\bf q$ dependence of the longitudinal form factor for the 12.5-20 MeV,  20-35 MeV  and 35-55 MeV regions in excitation energy.
%extracted from  Hotta74 and Goldman70\cite{Goldman70}  data compared to our parameterizations . 
%Since no data are available for the form factor for nuclear excitations the 36-55 MeV region in ${\rm ^{16}O}$ we assume that the form factor for this region in ${\rm ^{16}O}$ is similar to  the form factor for ${\rm ^{12}C}$ in this region.
}
\label{O16_from_Hotta}
\end{center}
\end{figure*}
%
  %% Table III  TTTTTTTTTTTTTTTTTTTTTTTTTTTTTTTTTTTTTTTTTTTTTTTTTTTTTTTTTTT
\begin{table*}
%\begin{table*}[ht]
\begin{center}
\begin{tabular}{|c|c||c|c|c||c|c|c||c|c|c||c|c|c||} \hline
State	&	          MeV	&   $N_1$  & $C_1$  & $\sigma_1$&$N_2$& $C_2$&$\sigma_2$&    $N_3$   & $C_3$ & $\sigma_3$&	 $a$	&       $b$           & Ref.	\\ \hline
$0_2^+ L$	&	6.0494	&	0.70	&	0.35	&	0.58	&	0.120&	1.20	&	0.500	&0.0050	&	4.30	&	1.60	&	0.220	&	6.00   &  Buti-86	\\
$3_1^-L$	&	6.1299	&	1.60	&	1.00	&	0.45	&	4.100&	1.07	&	1.700        &0.2000	&	2.55	&	2.40	&	3.100	&	1.10   &  Buti-86  \\
$2_1^+L$	&	6.9171	&	6.50	&	0.20	&	0.75	&	1.000&	1.10	&	0.955	&0.0032	&	5.30	&	1.35	&	5.000	&	2.50   &   Buti-86 	\\
$1_1^-L$	&	7.1169	&	0.95&	0.94	&	0.68	&	0.800&	1.50	&	1.200	&0.1000	&	2.40	&	1.55	&	0.400	&	3.00	  &  Buti-86   \\
%$1_1^-T$	&	7.1169	&		&		&		&	&		&		&		&		&		&		&	             &  Buti86   \\
$2_2^+L$	&	9.8445	&	0.10&	0.70	&	0.65	&	0.080&	1.70	&	1.200	&	0.0120	&	2.50	&	2.0	&	0.007	&	2.000    &	Buti-86 \\
$4_1^+L$	&	10.3560	&	0.09&	1.30	&	0.70	&	0.087&	2.05	&	1.200	&	0.0140	&	3.30	&	1.7	&	0.007	&	2.000    &	Buti-86 \\
$4_2^+L$	&	11.0967	&	0.04&	1.10	&	0.85 &	0.042&	2.20	&	1.100	&	0.0110	&	3.20	&	1.9	&	0.000	&	10.000  &	Buti-86 \\
$2_3^+L$	&	11.5200	&	2.00&	0.50	&	0.60	&	0.600&	1.00	&	1.050	&	0.0040	&	5.30	&	1.4	&	0.007	&	1.657   &  Buti-86	\\
$0_3^+L$	&	12.0490	&	1.15&	0.20	&	0.95	&	0.050&	1.50	&	0.850	&	0.0015	&	5.20	&	1.3	&	1.00	&	         4.00	  & Buti-86 \\ \hline
&               	12.5-20.0	&	2.00&	0.35	&	0.30	&	8.000&	0.0	&	1.90	&	 -  	&	-	&	-	&	0.007	&	1.66	& Hotta74\\ 
&  	                 20.0-35.0	&	18.00&	0.00&	0.40	&	7.000&	0.6	&	0.80	&	2.5000	&	1.0	&	1.5	&	-	&	-	& Hotta74\\ 
&	                35.0-55.0	&	0.8  &	0.00	&	0.30&	1.300&	0.6	&	3.00	&	-	&	-	&	-	&	0.004	&	1.66	& use $carbon$\\    
\hline
%
%$2_2^-T$	&	12.530&		&		&		&		&		&		&		&		&		&		&	&  Hotta74\\ 
%$1_4^-L$	&	13.2	&		&		&		&		&		&		&		&		&		&		&	& Bishop 63  \\ 
%$4_3^+T$	&	13.65	&		&		&		&		&		&		&		&		&		&		&	&  Bishop 63  \\ 
%\hline\hline
% 
%
 \hline
 \end{tabular}
\caption{
Parameterizations of the square of the  longitudinal (L) nuclear  excitation form factors in  ${\rm ^{16}O}$ (in units of $10^{-3}$).
Data taken from Buti-86\cite{Buti86} and Hotta74\cite{Hotta74}.  The parameterizations are functions of  $\bf q_{eff}^2$ in units of fm$^{-2}$.
}
\label{O16_states}
\end{center}
\end{table*} 
%  FFFFFFFFFFFFFFFFFF
%Fig. 11
\begin{figure*}
%\vspace{9pt}
\begin{center}
\includegraphics[width=3.5in, height=2.4in]{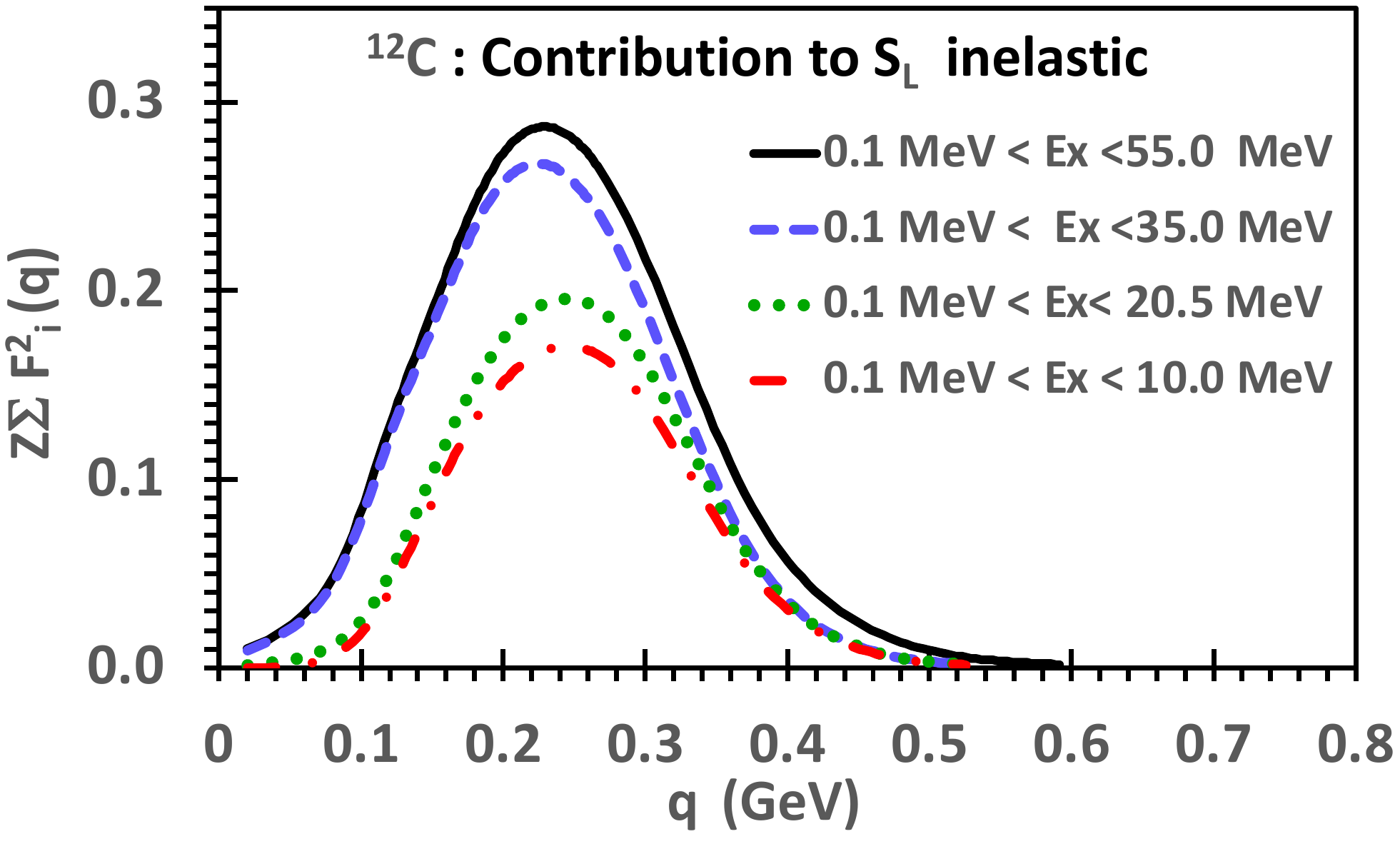}
\includegraphics[width=3.5in, height=2.4in]{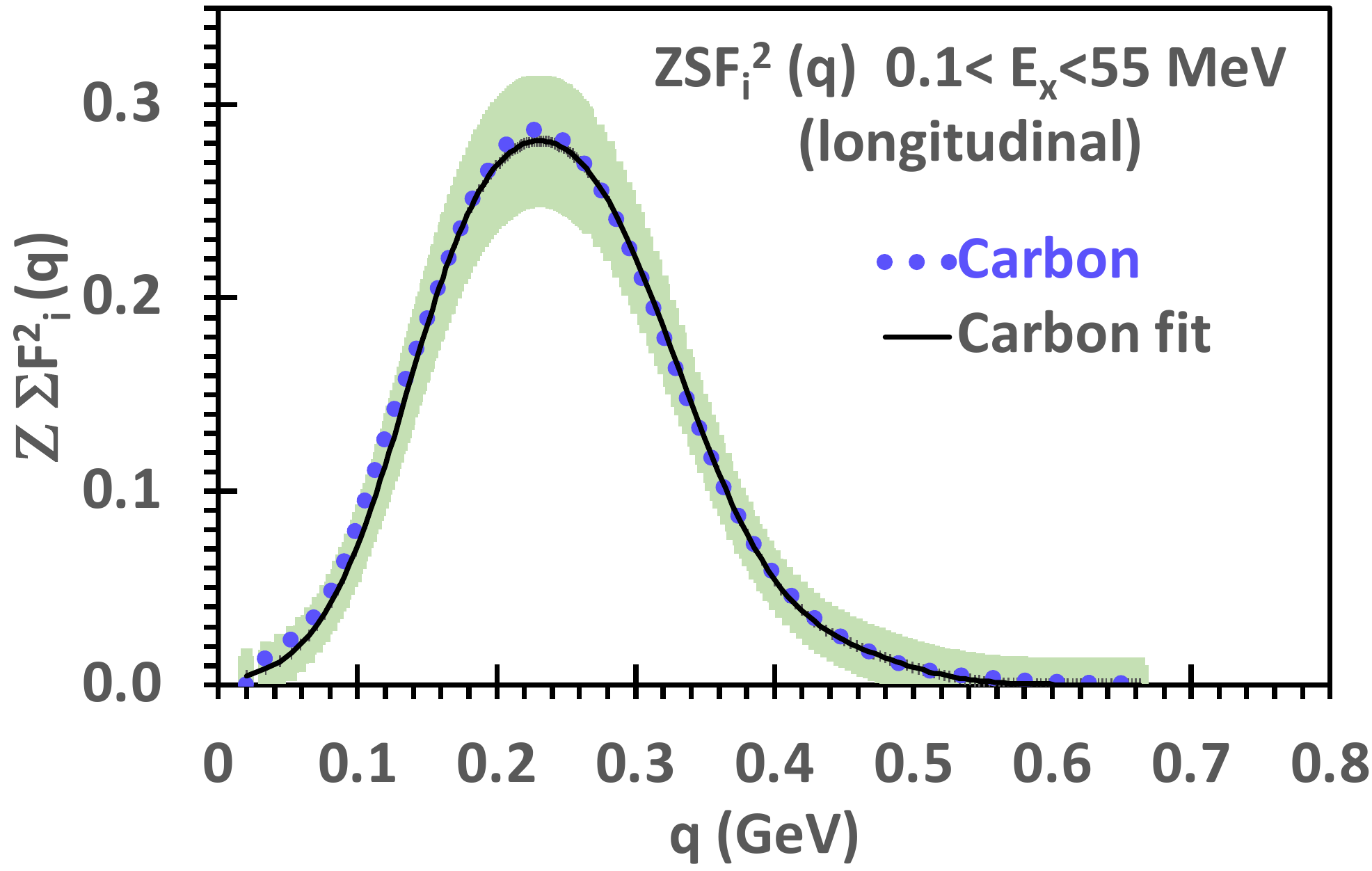}
\caption{{\bf Left panel}: 
  The contribution of longitudinal nuclear excitations (between 2 and 55 MeV) to $S_L({\bf q})$  in  ${\rm ^{12}C}$. {\bf Right panel}:  Our fit to the total contributions of all nuclear excitations below 55 MeV to  $S_L({\bf q})$ in ${\rm ^{12}C}$. The uncertainty is (shown as a green band) is 0.01 plus 10\% added in quadrature.
 }
\label{C12_States_LT}
\end{center}
\end{figure*}
%
%
%Fig 12
\begin{figure*}
%\vspace{9pt}
\begin{center}
\includegraphics[width=3.5in, height=2.4in]{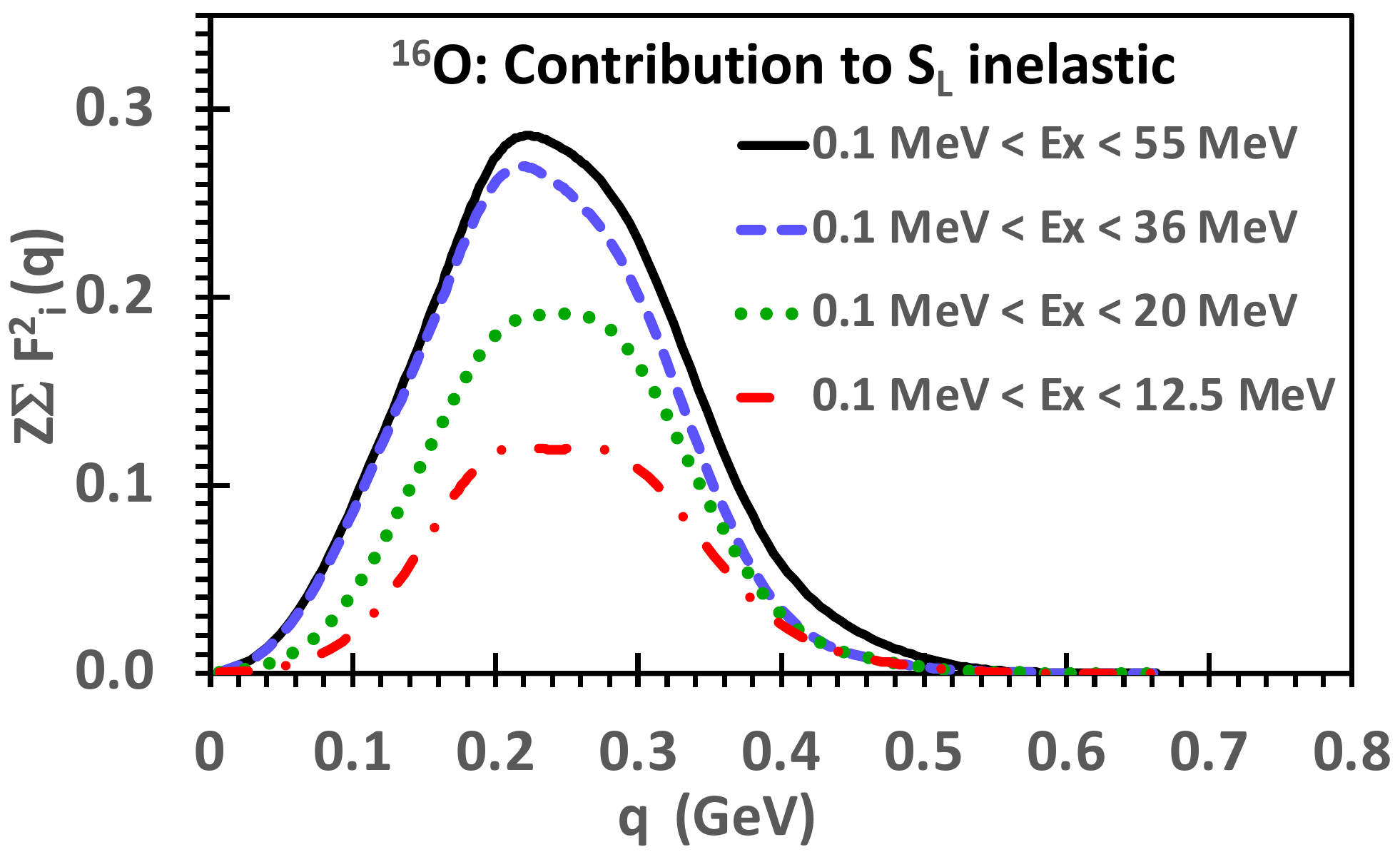}
\includegraphics[width=3.5in, height=2.4in]{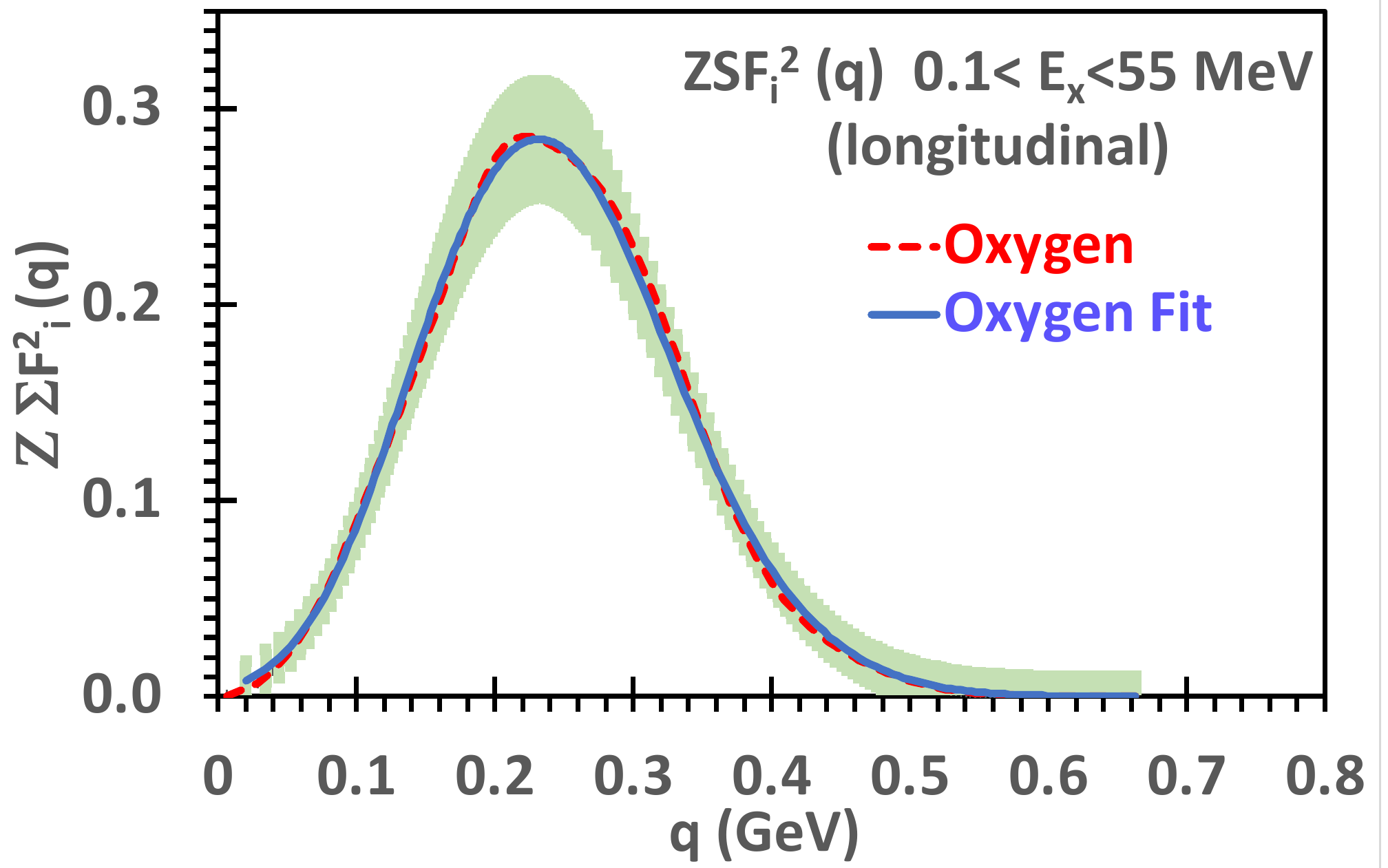}
\caption{ {\bf Left panel}: The contribution of longitudinal nuclear excitations (between 2 and 55 MeV) to $S_L({\bf q})$ in ${\rm ^{16}O}$. {\bf Right panel}: Our fit to the total contributions of all nuclear excitations below 55 MeV to $S_L({\bf q})$ for  ${ \rm ^{16}O}$. The uncertainty (shown as a green band) is 0.01 plus 10\% added in quadrature.
 }
\label{O16_States_LT}
\end{center}
\end{figure*}
%
% Fig 13
\begin{figure}
%\vspace{9pt}
\begin{center}
\includegraphics[width=3.5in, height=2.65 in]{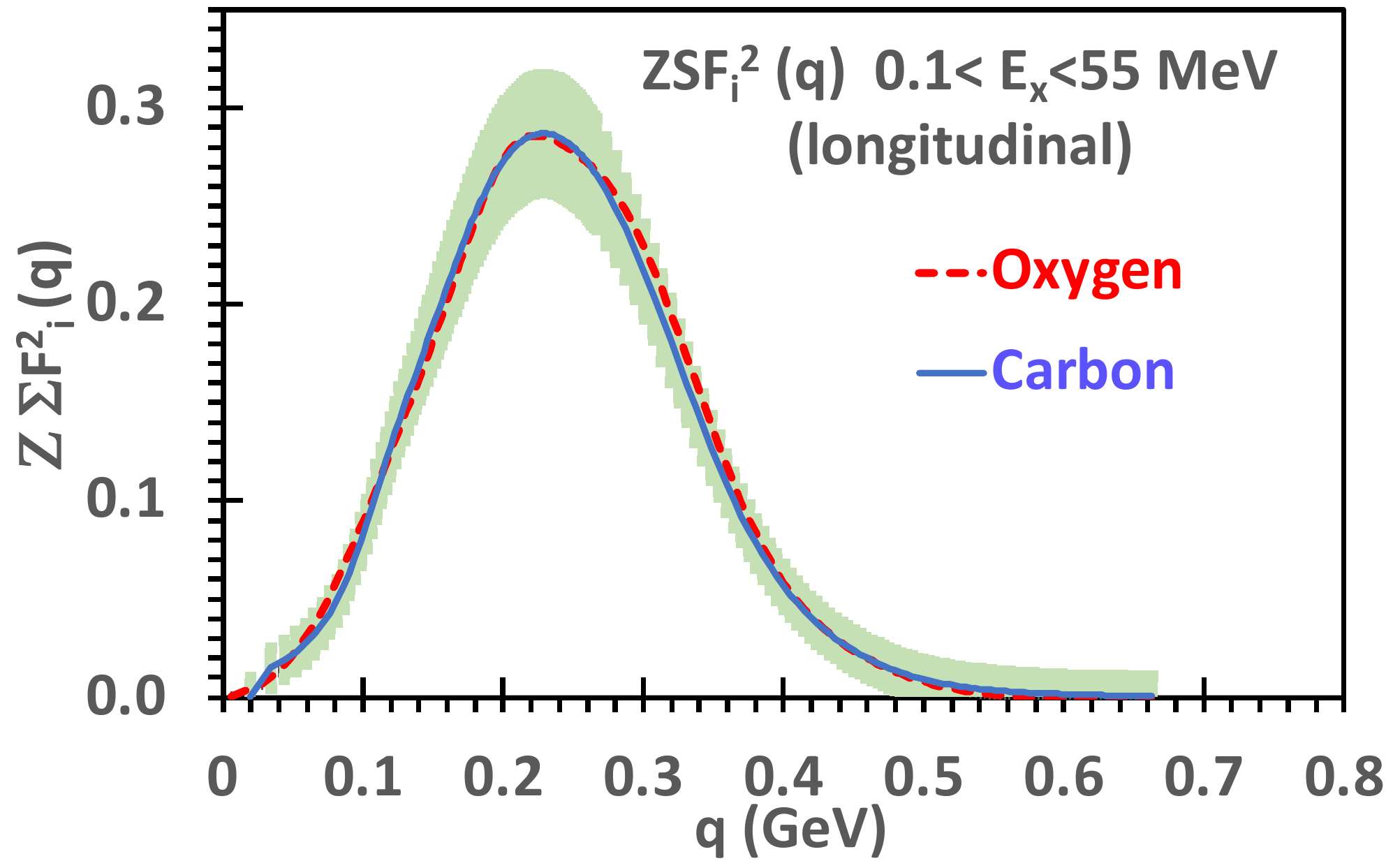}
\caption{A comparison of the contributions of nuclear excitations to $S_L({\bf q})$ in ${\rm ^{12}C}$ and ${\rm ^{16}O}$. The uncertainty (green band) in the total contribution of the excited states is  0.01 plus 10\% added in quadrature. 
} 
\label{C12_vs_O16}
\end{center}
\end{figure}
%
%  Section 5
 \section{Analysis of ${\rm ^{16}O}$ excited states}
\subsection{${\rm ^{16}O}$ excited states with $E_x<12.5 MeV$}
In order to minimuze correlations between our parameterizations of the form  factors for the nuclear excitations in ${\rm ^{12}C}$ and ${\rm ^{16}O}$ we parameterize the form factors for ${\rm ^{16}O}$ states using a somewhat different functional form. The form factors for the nuclear  excited states in  ${\rm ^{16}O}$  are  parameterized  as  $F_{iC}^2({\bf q^2})=Max(0.0,g_i^2)$ where
 %
 %Eq 27
  \begin{equation}
  \label{verses_q_O16}
g_i^2({\bf q_{eff}^2})= {\bf q_{eff}^2} \times \big[ \sum_{j=1}^{j=3}  N_j e^{-[({\bf q_{eff}^2}-C_j)/\sigma]^2} - a e^{-b{\bf q_{qeff}^2}}\big].
\end{equation}
Here,  $\bf q_{eff}^2$ is  in units of fm$^{-2}$.

The form factors for nuclear
excitations in ${\rm ^{16}O}$ with excitation energies below proton removal threshold (about 12 MeV) are easily measured because  there are is no contribution from QE scattering in this region. The nine longitudinal  form factors (squared) in  ${\rm ^{16}O}$  for excitation energies below 12.5 MeV are shown in  Fig. \ref{O16_Buti_states} on linear and logarithmic scales.  The data in the  figures are from Buti-86\cite{Buti86}. The solid  blue lines are our parameterizations using the parameters listed in Table \ref{O16_states}. 
%The parameters are given in Table \ref{O16_states}. 
%
\subsection{${\rm ^{16}O}$ excited states with $E_x>12.5 MeV$}
For excitation energy above 12.5 MeV there is a significant contribution from QE scattering. Here we group the states in two regions of excitation energy (12.5-20 MeV and 20-35 MeV).

The top row in Fig. \ref{O16_from_Hotta} shows the longitudinal response function $R_L({\bf q},E_x)$ for ${\rm ^{16}O}$  from Hotta74\cite{Hotta74} for three values of $\bf q$. The solid red line  is the original estimate of the QE contribution used in the  Hotta74 publication.  The solid  green line is the QE contribution determined using  the QE parameters from our universal fit to  all ${\rm ^{12}C}$ data. We find that the QE cross section predictions for ${\rm ^{16}O}$ using the parameters from the ${\rm ^{12}C}$ fit also describe all (but limited)  available data on ${\rm ^{16}O}$ as shown in \cite{short_letter}.  
We use the Hotta74 data to extract the longitudinal form factors for the nuclear excitation in ${\rm ^{16}O}$ in the 12.5-20 MeV and the 20-35 MeV
groupings in in excitation energy. 

The middle and bottom rows in Fig. \ref{O16_from_Hotta} show the extracted longitudinal form factor for the 12.5-20 MeV and 20-35 MeV groupings in excitation energy on linear (middle) and logarithmic (bottom) scales. Also shown is the form factor measurement from Goldman70\cite{Goldman70}.
%The form factors are extracted from the Hotta74  ${\rm ^{16}O}$ data.  
The longitudinal  form factor measured by Goldman70 for the  20--30 MeV grouping in excitation energy has been corrected by subtracting the QE contribution  (from our universal fit) and extending the excitation range to 20-35 MeV. 

Since no data are available for the form factor for nuclear excitations the 36-55 MeV region in ${\rm ^{16}O}$ we assume that the form factor for ${\rm ^{16}O}$ is the same as the form factor for ${\rm ^{12}C}$ in this region.
%
%  Fig.  14
\begin{figure}[ht]
%\vspace{9pt}
\begin{center}
\includegraphics[width=3.4in, height=2.5in] {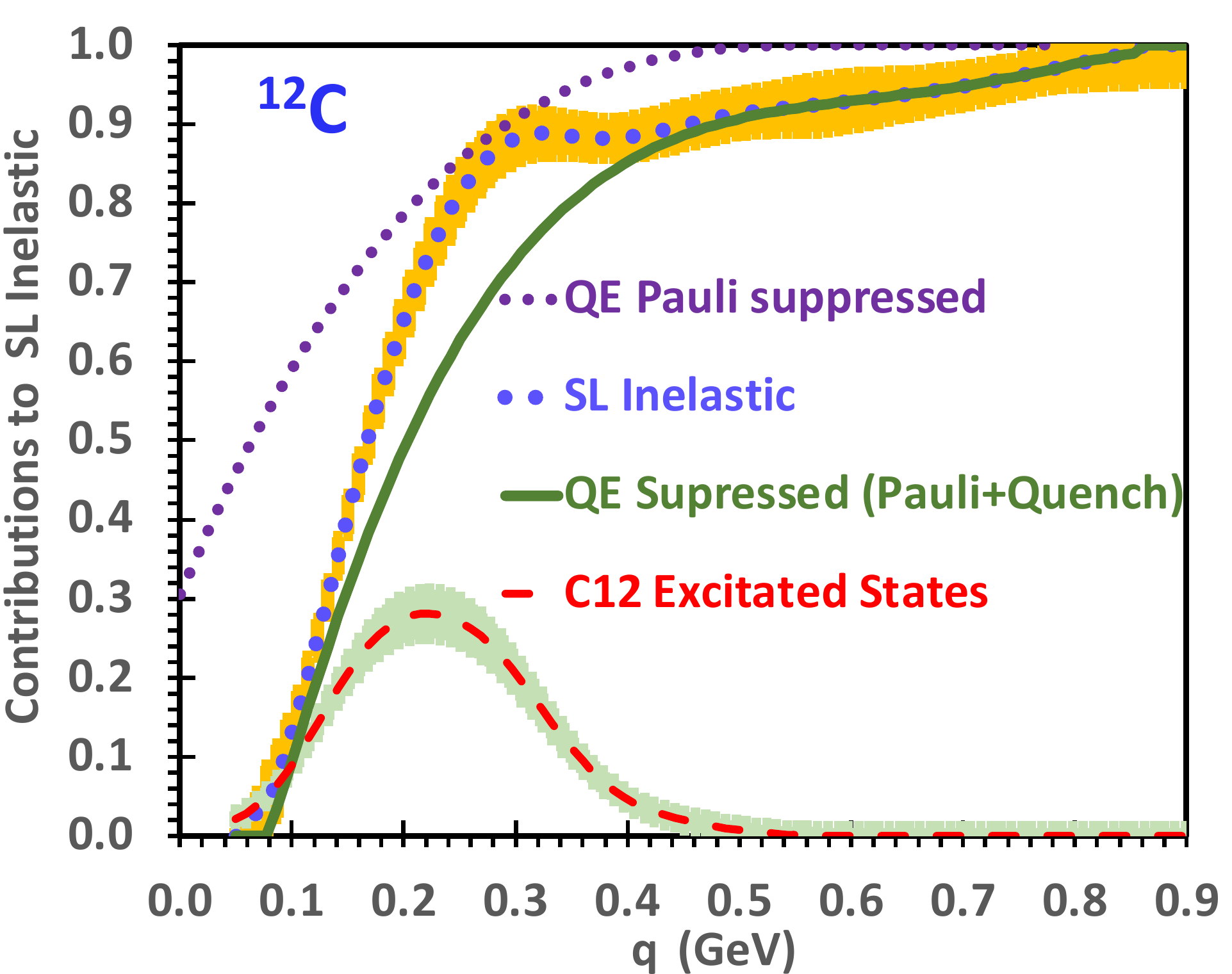}
\caption { The various contributions\cite{short_letter} to $S_L({\bf q})$ 
%the measured  "Inelastic Coulomb Sum Rule" 
 for ${\rm ^{12}C}$ (dotted blue with yellow error band) including:  QE  with Pauli suppression only (dotted-purple), QE suppressed by both "Pauli" and   "Longitudinal Quenching" (solid-green), and the contribution of  nuclear excitations (red-dashed with green error band).
}
\label{SL_components}
\end{center}
\end{figure}
%
%Fig. 15
\begin{figure*}
%\vspace{9pt}
\begin{center}
\includegraphics[width=3.4in, height=2.5in] {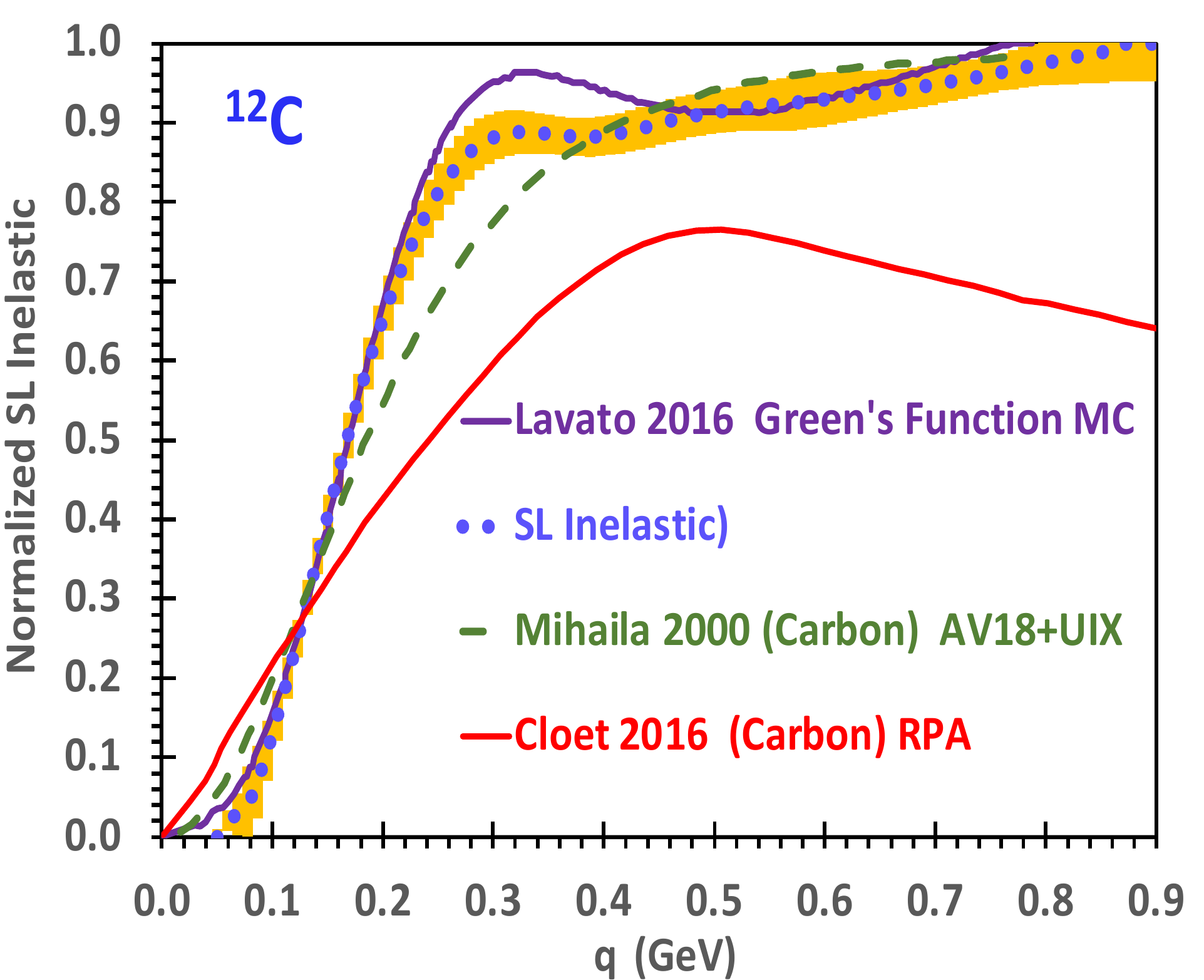}
\includegraphics[width=3.4in, height=2.5in] {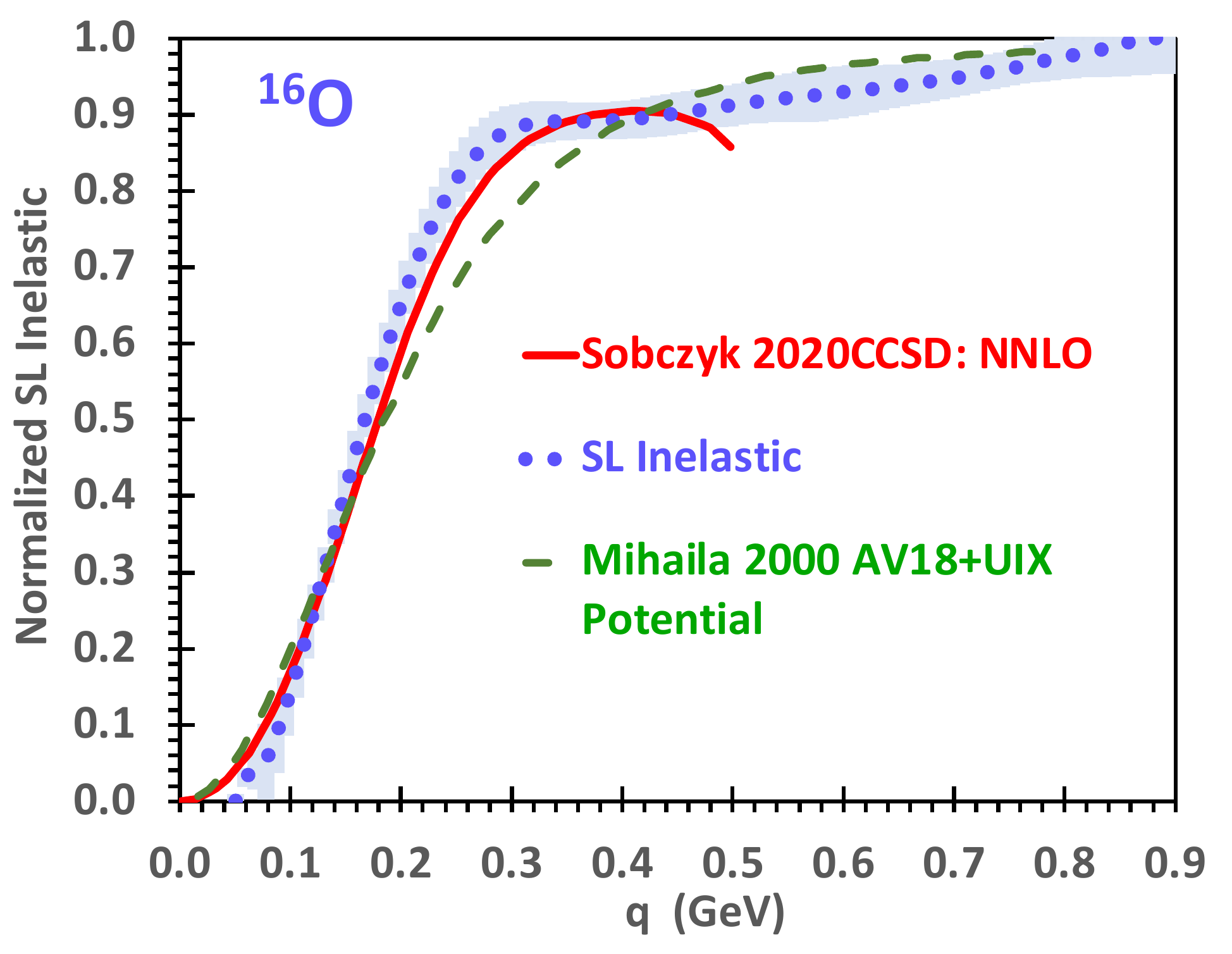}
\caption{   
{\bf Left panel}:  $S_L({\bf q})$ for $\carbon$ (dotted-blue with yellow error band) compared to theoretical calculations including  Lovato 2016 \cite{Lovato2016} (solid-purple), (Mihaila 2000\cite{microscopic} (dashed-green), and  RPA Cloet 2016\cite{Cloet}  (solid-red). {\bf Right panel}: $S_L({\bf q})$ for $\oxygen$  (dotted-black with green error band) compared to theoretical  calculations of  Sobczyk 2020\cite{Coupled}  (red-dashed) and Mihaila 2000 (dotted-dashed).}
\label{SL_for_C12_O16}
\end{center}
\end{figure*}

\section{Contribution of Nuclear Excitations to $S_L({\bf q})$ in ${\rm ^{12}C}$ and ${\rm ^{16}O}$ }
The contributions of nuclear excitation to $S_L({\bf q})$ (Eq. \ref{SLINE}) in ${\rm ^{12}C}$ and ${\rm ^{16}O}$ are  calculated using the form factor parameterizations given in  Tables \ref{excited_states1}, \ref{excited_states2} and \ref{O16_states}. The left side panels of  Figures  \ref{C12_States_LT} and \ref{O16_States_LT} show the contributions  of nuclear excitations (with excitation energies below 10 MeV, 20.5 MeV, 30 MeV and 55 MeV) to  $S_L({\bf q})$  for  ${\rm ^{12}C}$ and ${\rm ^{16}O}$, respectively.  
  
The  total contribution to $S_L({\bf q})$ can be  parametrized as follows: 
% Eq 12
%=NORM*EXP(-1*(GC2-CENTER)^2/SIGMA^2)+NORM2*EXP(-1*(GC2-CENTER2)^2/SIGMA2^2)
\begin{eqnarray}
\label{all states}
Z \sum_{all}^{L} F^2_i({\bf q}) &=&N_1 exp((x-C_1)^2/D_1^2)\nonumber\\
&+& N_2 exp(-(x-C_2)^2/D_2^2) \nonumber\\
&+& N_3 exp(-(x-C_3)^2/D_3^2)
\end{eqnarray}
where  x= ${\bf q}/K_F$. For ${\rm ^{12}C}$  $K_F$=0.228 GeV,  $N_1$= 0.260, $C_1$=1.11,  $D_1$=0.50, $N_2$= 0.075, $C_2$=0.730,  $D_2$=0.30, and  $N_3$= 0.01, $C_3$=2.0,  $D_3$=0.30.
The fit and the data are shown on the right side panel of Fig.  \ref{C12_States_LT}.

For ${\rm ^{16}O}$   $K_F$=0.228 GeV,  $N_1$= 0.240, $C_1$=1.07,  $D_1$=0.48, $N_2$= 0.073 $C_2$=0.70,  $D_2$=0.37, and 
$N_3$= 0.039, $C_3$=1.55,  $D_3$=0.50.
The fit and the data are shown on the right side panel of Fig.  \ref{O16_States_LT}.

Fig. \ref{C12_vs_O16}  shows a comparison of the contributions of all excited states to the $S_L({\bf q})$ for ${\rm ^{12}C}$ and ${\rm ^{16}O}$. The uncertainty in the total contribution of the excited states for both nuclei  is 0.01 plus 10\% added in quadrature. These data indicate that the contribution of  nuclear excitations to $S_L({\bf q})$  in ${\rm ^{16}O}$ is consistent with being equal to the contribution of the nuclear excitations in ${\rm ^{12}C}$ within errors.
 The total contribution of all states with excitation energy below 55 MeV is largest at  $\bf q$=0.22 GeV, where it reaches a maximum of 0.29$\pm$ 0.03.

%  Eq 3
 %\begin{eqnarray}
 %F_{oC}^2({\bf q}) &=& G_E^{2} (Q^2)\times  F_{o}^2({\bf q})\\
%F_{iC}^2({\bf q}) &=& G_E^{2} (Q^2)\times  F_{i}^2({\bf q}) \nonumber
%F_{iC}^2(\{bf fq}) &=& G^{\prime 2} \times  F_{i}^2(\{bf fq})\nonumber
%\end{eqnarray}
%
\section{Updated extraction of $S_L({\bf q})$ for ${\rm ^{12}C}$ and ${\rm ^{16}O}$  }
In our previous paper\cite{short_letter} 
we performed a fit to all electron scattering data on ${\rm ^{12}C}$ and ${\rm ^{16}O}$. 
We found that the QE transverse response function is enhanced at intermediate ${\bf q}$ and the  longitudinal response function is quenched at low ${\bf q}$. We used the fits in combination with the fits to nuclear excitations to extract $S_L({\bf q})$ for ${\rm ^{12}C}$ and ${\rm ^{16}O}$.   In our previous paper we used a very conservative estimate of   the uncertainty in the total contribution of the excited states (0.01 plus 15\% added in quadrature).  In this paper we have updated our fits to the form factors for individual nuclear excitations. We find that  the updated total contribution of nuclear excitations to  $S_L({\bf q})$ for ${\rm ^{12}C}$ and ${\rm ^{16}O}$ is unchanged, but  a smaller conservative estimate  (0.01 plus 10\% added in quadrature) is more appropriate.

Fig.~\ref{SL_components}  shows the various contributions to the extracted $S_L({\bf q})$  for ${\rm ^{12}C}$  (dotted blue line with yellow error band). Shown are the QE contribution with only Pauli suppression (dotted-purple), the QE contribution  suppressed by both "Pauli Suppression" and the  longitudinal quenching factor  $F^L_{quench}({\bf q})$  labeled as QE suppressed (Pauli+Quench) (solid-green),  and  the  contribution of nuclear excitations (red dashed line).

The left panel of Fig. \ref{SL_for_C12_O16} shows a  comparison of the extracted $S_L({\bf q})$ for $\carbon$ (dotted-blue curve with yellow error band) to theoretical calculations.  These include the Lovato 2016\cite{Lovato2016} "First Principle Green's Function Monte Carlo" (GFMC) calculation (solid-purple line),  Mihaila\cite{microscopic}  2000   Coupled-Clusters based calculation (AV18+UIX potential, dashed-green), and Cloet 2016\cite{Cloet} RPA calculation (RPA solid-red). Our measurement for  $\carbon$ are in disagreement with Cloet 2016 RPA, and in reasonable agreement with  Lovato 2016 except near $\bf{q}\approx$ 0.30 GeV where the contribution from nuclear excitations is significant.
  
The right panel  of Fig. \ref{SL_for_C12_O16} shows $S_L({\bf q})$ for $\oxygen$  (dotted-blue  with green error band) compared to theoretical calculations.
   %for $\oxygen$. 
   These include the Sobczyk  2020\cite{Coupled} "Coupled-Cluster with Singles-and Doubles (CCSD) NNLO$_{sat}$" (red-dashed line),  and   Mihaila 2000\cite{microscopic}   Coupled-Cluster  calculation with (AV18+UIX potential, dashed green line). The data are in reasonable agreement with Sobczyk  2020.
%section 9
   \section{Summary}
   We report on empirical parameterizations of longitudinal and transverse nuclear excitation electromagnetic form factors in ${\rm ^{12}C}$ and ${\rm ^{16}O}$ and extract the contribution of  nuclear excitations to the Normalized Inelastic Sum Rule $S_L({\bf q})$ as a function momentum transfer $\bf q$.  We find that the total contribution is significant (0.29$\pm$0.030) at $\bf q$= 0.22 GeV. The  total contributions of nuclear excitations in ${\rm ^{12}C}$ and ${\rm ^{16}O}$ are consistent with being equal within errors.
 Since the cross sections for nuclear excitations are significant at  low $\bf q$, the radiative tails from nuclear excitations should be included in precise calculations of radiative corrections to quasielastic electron scattering at  low $\bf q$ and deep-inelastic electron scattering at large $\nu$.
   
 The parameterization also serves as a benchmark in testing theoretical modeling of electron and neutrino scattering cross sections at low energies. Theoretical studies of the  excitation of nuclear states in electron and  neutrino scattering\cite{Pandey1,Pandey2,Pandey3} indicate that  both are equally significant  at low values of $\bf q$.  Therefore, nuclear excitations should be included in both electron and neutrino MC generators.  We note that for excitation energies above proton removal threshold (about 16 MeV in ${\rm ^{12}C}$ and 12 MeV in ${\rm ^{16}O}$) the decays of  nuclear excitations can have a proton in the final state and  therefore cannot be distinguished  experimentally from QE scattering in low resolution neutrino experiments. 
  \section{Acknowledgements}
 Research supported by the U.S. Department of Energy under University of Rochester grant number DE-SC0008475,  and the Office of Science, Office of Nuclear Physics under contract DE-AC05-06OR23177.
 %
%sec
%\newpage
%    REFERNCES  -------------------------------------------------------------------------------
%

%

\begin{thebibliography}{9}
% ref 1
 \bibitem{CSR}  D. Drechsel and M M Giannini 1989 Rep. Prog. Phys. 52 1083 (eq. 7.9); T. de Forest Jr. and J.D. Walecka, Advances in Physics, 15:57, 1-109 (1966) (eq. 6.8).
 %  ref 2
 \bibitem{short_letter} A. Bodek and M. E. Christy "Extraction of the Coulomb Sum Rule, Transverse Enhancement, and Longitudinal Quenching from an Analysis of all Available e-12C and e-16O Cross Section Data", arXiv:2208.14772 (to be  published in Phys. Rev. C letters 2022)
 %ref 3
 \bibitem{Lovato2016} A. Lovato et. al, Phys. Rev. Lett. 117, 082501 (2016)
% ref 4
% Mihaila (2000)) 
\bibitem{microscopic}    Bogdan Mihaila and Jochen H. Heisenberg, Phys, Rev. Lett.  84 (2000) 1403.
2009.01761 [nucl-th]
%
% Lonardoni  (2018
%\bibitem{AFDMC} D. Lonardoni, J. Carlson, S. Gandolfi, J.E. Lynn, K.E. Schmidt, A. Schwenk, and X.B.Wang, 
% Properties of Nuclei up to A = 16 using Local Chiral Interactions, 
%Phys. Rev. Lett. 120, 122502 (2018);   D. Lonardoni, S. Gandolfi, J. E. Lynn, C. Petrie, J. Carlson, K.E. Schmidt, and A. Schwenk,
%  Auxiliary field diffusionMonte Carlo calculations of light and medium-mass nuclei with local chiral interactions,
 %Phys. Rev. C 97, 044318 (2018).  53
%Mihaila et al. [18] AV18+UIX potential)
%    Mihaila-Heisenberg(2000)
%
 % ref 5
%Sobczyk (2020) 
\bibitem{Coupled}   J. E. Sobczyk, B. Acharya, S. Bacca, and G. Hagen  Phys.Rev.C 102 (2020) 064312.
%(arXiv: 2009.01761 [nucl-th])
% Ref 6
\bibitem{Pandey1} V. Pandey, N. Jachowicz, T. Van Cuyck, J. Ryckebusch,  and M. Martini, Phys. Rev. C 92, 024606 (2015), %(arXiv:1412.4624 [nucl-th])
% ref 7
\bibitem{Pandey2} M. Martini, N. Jachowicz, M. Ericson, V. Pandey, T. Van Cuyck, and N. Van Dessel, Phys. Rev. C 94, 015501 (2016).
% ref 8
\bibitem{Pandey3} V. Pandey, N. Jachowicz, M. Martini, R. Gonzalez-Jimenez, J. Ryckebusch, T. Van Cuyck, and N. Van Dessel, Phys. Rev. C 94, 054609 (2016).
%  Ref 9
\bibitem{Crannell1} H. L. Crannell and T. A. Griffy, Phys. Rev. 136, B1580 (1964); H. Crannell, Phys. Rev. 148, 1107 (1966).
 % Rev 10
\bibitem{Yamaguchi} Y. Yamaguchi et al., Phys. Rev. D3, 1750 (1971).
   % ref 11
   \bibitem{ledex_exp} P. Gueye et al., Eur. Phys. J. A 56, 126 (2020).
   %(https://doi.org/10.1140/epja/s10050-020-00135-7) (arXiv:1805.12441) 
%ref 12
\bibitem{GEp} J.A. Caballero,  M. C. Martinez, J. L. Herraiz, J.M. Udias  Physics Letters B 688. 250 (2010).
% % Ref 13
\bibitem{product}   P. Gueye et.al.	Eur. Phys. J. A 56, 126 (2020);   I. Sick, Nuclear Physics A218, 509 (1974).
%  ref 14
 \bibitem{BBBA} A. Bodek, S. Avvakumov, R. Bradford, H. Budd,  Eur. Phys. J C53, 349 (2008).
 %  Ref 15
  \bibitem{q_effective}  T. W. Donnelly and D. Walecka, Annu. Rev. Nucl. Sci. 1975.25:329. 
% Ref  16
 \bibitem{Hofstadter}  R. Hofstadter, Annu. Rev. Nucl. Sci. 1957.7:231-316; ibid Rev. Modern Physics, 28, 214 (1956);
 J. Fregeau and R. Hofstadter, Phys Rev. 99.1503 (1955).
  %
   %ref 17
   %\cite{C12_FF}
 \bibitem{C12_FF} Carbon  Elastic and 4.43 MeV and 9.65 MeV form factor  measurements: Jerome H. Fregeau, Phys. Rev. 104, 225 (1956) (80, 150 MeV); H. L. Crannell and T. A. Grippy, Phys. Rev.136, B1580 (1964) (187, 250,  and 300 MeV) ;F. E. Eherenberg et. al., Phys. Rev. 113, 666 (1959) (420 MeV);  H. Crannell, Phys. Rev. 148, 1107 (1966) (600 and  800 MeV);  I. Sick and J. McCarthy, Nucl. Phys. A 150, 631 (1970) (374.5 and 747.2 MeV, elastic only).  
% Ref 18
    %  Chernykh:2010\cite{FF_765}
   \bibitem{FF_765} M. Chernykh et.al. Phys. Rev. Lett. 105:022501,2010. 
%(arXiv:1004.3877 [nucl-ex]).
   %Ref. 19
   \bibitem{FF_1084}Y. Torizuka et. al., Phys. Rev. Lett. 22, 544(1969), M. C. A. Campos et. el. , Phys. Lett. B349, 433 (1995).
 %Ref 20
 \bibitem{FF_1271_1511}
 J.B. Flanz, R. S. Hicks, R. A. Lindgren, G. A. Peterson, J. Dubach and C. Haxton, Phys. Rev. Lett. 43,  1923 (1979).
 %%Ref 21 
\bibitem{FF_1408} N. Nakada, Y. Torizuka and H. Horikawa, Phys. Rev. Lett. 27, 795 (1971).
% Ref 22
 \bibitem{Hicks84} R. S, Hicks et. al. Phys. ReV.C30, 1 (1984).
 %Ref 23
 \bibitem{superscaling} C. Maieron, T.W. Donnelly, I. Sick,  Phys.Rev. C65, 025502, (2002);  J.E. Amaro, M.B. Barbaro, J.A. Caballero,  T.W. Donnelly, A. Molinari, and I. Sick,  Phys. Rev. C 71, 015501 (2005); J.E. Amaro, M.B. Barbaro, J.A. Caballero, R. Gonzalez-Jimenez, G.D. Megias, I. Ruiz Simo, J. Phys. G: Nucl. Part. Phys. 47, 124001 (2020).
%Ref 24
 \bibitem{Rosenfelder}  R. Rosenfelder, Ann. Phys. 128, 188 (1980);   G. D.  Megias,  M. V. Ivanov, R. Gonzalez-Jimenez, M. B. Barbaro,J. A. Caballero, T. W. Donnelly,  J. M. Udias, Phys. Rev. D 89, 093002 (2014); G.D. Megias Vazquez (Tesis Doctoral). Universidad de Sevilla, Sevilla (2017).
 % ref 25
 %54 MeV  180 deg
 \bibitem{Goldemberg64} J. Goldemberg and W. C. Barber, Phys. Rev. 134,  B963 (1964).
%ref 26
%65 MeV 180 deg  deForest (1965)
\bibitem{deForest65} T. de Forest, J. D. Walecka, G. Vanpraet, and W. C. Barber, Phys. Letters 16, 311  (1965). 
%% ref 27
 \bibitem{Buti86}T. N. Buti et al., Phys. Rev., C33, 755 (1986).
 % ref 28
  \bibitem{Hotta74}T. A.  Hotta,  K. Itoh and T. Saito., Phys. Rev. Lett. 33, 790 (1974).
  %ref 29
 \bibitem{Goldman70} A. Goldmann and M. Stroetzel, Z. Phys. 239, 235 (1970).
%    
\bibitem{Cloet} Ian C. Cloet, Wolfgang Bentz, Anthony W. Thomas, Phys. Rev. Lett. 116, 032701 (2016).
\end{thebibliography}
\end{document}